\documentclass[twocolumn]{IEEEtran}
\usepackage{times,amssymb,amsmath,amsfonts,float,color,cite,bbm,mathrsfs,float,stmaryrd}
\usepackage{latexsym}
\usepackage{theorem}
\usepackage{graphicx}
\usepackage{subcaption}
\usepackage{enumitem}
\usepackage{algorithm,algpseudocode}
\usepackage[normalem]{ulem}

\newtheorem{theorem}{Theorem}[section]
\newtheorem{lemma}[theorem]{Lemma}

\newtheorem{proposition}[theorem]{Proposition}

\newtheorem{corollary}[theorem]{Corollary}
\newtheorem{definition}[theorem]{Definition}

\newtheorem{example}[theorem]{Example}
\newtheorem{remark}[theorem]{Remark}
\newtheorem{construction}[theorem]{Construction}

\renewcommand{\thefigure}{\thesection.\arabic{figure}}
\usepackage{amsmath}

\allowdisplaybreaks

\renewcommand{\IEEEQED}{\hfill \ensuremath{\Box}}

\makeatletter
\renewcommand{\@endtheorem}{\endtrivlist}
\makeatother

\newcommand\remove[1]{}

\makeatletter
\renewcommand{\thefigure}{{\@arabic\c@figure}}
\renewcommand{\fnum@figure}{{\bf Figure\,\thefigure}}
\makeatother

\def\mathbi#1{{\textbf{\textit #1}}}
\newcommand\nc\newcommand
\nc{\cA}{\mathcal{A}}\nc{\cB}{\mathcal{B}}\nc{\cC}{\mathcal{C}}\nc{\cD}{\mathcal{D}}
\nc{\cE}{\mathcal{E}}\nc{\cF}{\mathcal{F}}\nc{\cG}{\mathcal{G}}\nc{\cH}{\mathcal{H}}
\nc{\cI}{\mathcal{I}}\nc{\cJ}{\mathcal{J}}\nc{\cK}{\mathcal{K}}\nc{\cL}{\mathcal{L}}
\nc{\cM}{\mathcal{M}}\nc{\cN}{\mathcal{N}}\nc{\cO}{\mathcal{O}}\nc{\cP}{\mathcal{P}}
\nc{\cQ}{\mathcal{Q}}\nc{\cR}{\mathcal{R}}\nc{\cS}{\mathscr{S}}\nc{\cT}{\mathcal{T}}
\nc{\cU}{\mathcal{U}}\nc{\cV}{\mathcal{V}}\nc{\cW}{\mathcal{W}}\nc{\cX}{\mathcal{X}}
\nc{\cY}{\mathcal{Y}}\nc{\cZ}{\mathcal{Z}}

\nc{\bba}{\mathbf{a}}\nc{\bbb}{\mathbf{b}}\nc{\bbc}{\mathbf{c}}\nc{\bbd}{\mathbf{d}}
\nc{\bbe}{\mathbf{e}}\nc{\bbf}{\mathbf{f}}\nc{\bbg}{\mathbf{g}}\nc{\bbh}{\mathbf{h}}
\nc{\bbi}{\mathbf{i}}\nc{\bbj}{\mathbf{j}}\nc{\bbk}{\mathbf{k}}\nc{\bbl}{\mathbf{l}}
\nc{\bbm}{\mathbf{m}}\nc{\bbn}{\mathbf{n}}\nc{\bbo}{\mathbf{o}}\nc{\bbp}{\mathbf{p}}
\nc{\bbq}{\mathbf{q}}\nc{\bbr}{\mathbf{r}}\nc{\bbs}{\boldsymbol{S}}\nc{\bbt}{\mathbf{t}}
\nc{\bbu}{\mathbf{u}}\nc{\bbv}{\mathbf{v}}\nc{\bbw}{\mathbf{w}}\nc{\bfx}{\boldsymbol{x}}
\nc{\bby}{\mathbf{y}}\nc{\bbz}{\mathbf{z}}

\newcommand{\mathset}[1]{\left\{#1\right\}}
\newcommand{\abs}[1]{\left|#1\right|}
\newcommand{\ceilenv}[1]{\left\lceil #1 \right\rceil}
\newcommand{\floorenv}[1]{\left\lfloor #1 \right\rfloor}
\newcommand{\parenv}[1]{\left( #1 \right)}
\newcommand{\sparenv}[1]{\left[ #1 \right]}

\nc{\set}[1]{\llbracket #1 \rrbracket}

\newcommand{\bal}[1]{\begin{align}\label{#1}}
\newcommand{\eal}{\end{align}}

\renewcommand{\le}{\leqslant}
\renewcommand{\leq}{\leqslant}
\renewcommand{\ge}{\geqslant}
\renewcommand{\geq}{\geqslant}

\renewcommand{\Bbb}{\mathbb}

\newcommand{\Cref}[1]{Co\-ro\-lla\-ry\,\ref{#1}}

\renewcommand{\Bbb}{\mathbb}

\newcommand{\N}{{\Bbb N}}
\newcommand{\Q}{{\Bbb Q}}
\newcommand{\R}{{\Bbb R}}
\newcommand{\Z}{{\Bbb Z}}

\DeclareMathOperator{\db}{DB}
\DeclareMathOperator{\Tr}{Tr}
\DeclareMathOperator{\ccap}{cap}
\DeclareMathOperator{\rank}{rk}

\DeclareMathOperator{\per}{\cP}

\outer\def\proclaim #1. #2\par{\medbreak
 \noindent{\bf#1.\enspace}{\sl#2\par}%
 \ifdim\lastskip<\medskipamount \removelastskip\penalty55\medskip\fi}

\begin{document}

\title{{Recoverable Systems}}

\author{\IEEEauthorblockN{Ohad Elishco} \and\hspace*{.5in} \IEEEauthorblockN{Alexander Barg} 
}

\maketitle

{\renewcommand{\thefootnote}{}\footnotetext{

\vspace{-.2in}
 
\noindent\rule{1.5in}{.4pt}

{ This work was presented in part at the 2021 IEEE International Symposium on Information Theory.

Ohad Elishco was with ISR, University of Maryland, College Park, MD 20742, USA. He is now with School of EE, Ben-Gurion University of the Negev, 
Beer Sheva, Israel. Email: elishco@gmail.com. This research was supported in part by the NSF-BSF grant 2020762.

Alexander Barg is with ISR and Department of ECE, University of Maryland, College Park, MD 20742, USA, and also with Inst. Inform. Trans. Probl., Russian Academy of Sciences, 127051 Moscow, Russia. Email abarg@umd.edu. This research was supported in part by the NSF grants CCF1814487, CCF2104489, as well as by NSF-BSF grant CCF-2110113.
}
}}


\begin{abstract}
Motivated by the established notion of storage codes, we consider sets of infinite sequences over a finite alphabet such that every $k$-tuple of consecutive entries is uniquely recoverable from its $l$-neighborhood in the sequence. We address the problem of finding the maximum growth rate of the set, which we term capacity, as well as constructions of explicit families that approach the optimal rate. The techniques that we employ rely on the connection of this problem with constrained systems. 
In the second part of the paper we consider a modification of the problem wherein the entries in the sequence are viewed as random variables over a finite alphabet that follow some joint distribution, and the recovery condition requires that the Shannon entropy of the $k$-tuple conditioned on its $l$-neighborhood be bounded above by some $\epsilon>0.$ We study properties of measures on infinite sequences that maximize the metric entropy under the recoverability condition. Drawing on tools from ergodic theory, we prove some properties of entropy-maximizing measures. We also suggest a procedure of constructing an $\epsilon$-recoverable measure from a corresponding deterministic system.
\end{abstract}


\section{Introduction}
Consider the set of binary sequences $X_n\subset\{0,1\}^n$ with the property that every bit in the sequence is uniquely determined by its neighbors. 
What is the growth rate of the maximum size of $X_n$ with this property as $n$ increases without limit? 
Moreover, what can be said about the growth rate if we permit ourselves recovery with high probability rather than a deterministic decision? 
These questions form the starting point of this study which has its motivation in the recently established areas of storage coding and index coding, and draws on 
connections with constrained systems and entropy theory with the goal of establishing various ``capacity'' results for $X_n$ and associated 
concepts. 

Our main object is a {\em recoverable system} over a finite alphabet, which we proceed to define. We consider the set $Q^\Z$ of bi-infinite sequences 
$x=(\dots,x_{-2},x_{-1},x_0,x_1,x_2,\dots)$ over a finite alphabet $Q$. A set $X\subset 
Q^{\Z}$ is called shift invariant if $TX=X,$ where $T$ is a shift by one position to the left. Let $[k]:=\{0,1,\dots,k-1\}$ and define the $l$-neighborhood $N_l([k])$ of the subset of coordinates $[k]$ as a set of all indices $j\in\Z\setminus [k]$ such that $|j-i|\le l$ for some $i\in[k]$. The elements of the set $N_l([k])$ have a natural ordering, and we always think of the neighborhood as an ordered set.

\begin{definition}\label{def:RS} Let $Q$ be a finite alphabet and let $k,l\in\N$ be two integers. A set $X\subset 
Q^{\Z}$ is called a $(k,l)$-\textbf{recoverable system} if $X$ is shift invariant and if there exists a 
deterministic function $f: Q^{2l}\to Q^k$ such that for 
every $x\in X$, 
   $$
   f(x_{N_l([k])})=x_{[k]}.
   $$
\end{definition}
In other words, we look at sequences on $\Z$ (viewed as a graph) such that the value of a set of $k$ consecutive vertices is recoverable from their $l$-neighborhood. The assumption of shift invariance is natural because the recovery property does not depend on the location of the $k$-tuple in $x$. 
For this reason it suffices to define the recovery function only for the set $x_{[k]}.$

A similar notion was studied earlier for general finite graphs under the name of {\em 
storage codes}, defined in \cite{Maz2015,MazMcgVor2019}, and in \cite{Riis2007,Cameron2016,Gadouleau2018} in the context of {\em guessing games} and dynamical systems on graphs. It has also been shown in \cite{AloLubStaWeiHas2008, Maz2015} that the capacity problem of storage codes on graphs is close (and for a sufficiently large alphabet is equivalent) to the minimum attainable rate of {\em index coding} as well as to the success probability in guessing games. In this context the concept of storage 
codes was introduced and studied in  \cite{AK2015},\cite{ShanDim2014}. 

An important difference between the storage codes and the above definition is that  a storage code allows its own recovery function for every vertex while a recoverable system relies on a single repair rule for every vertex. Arguably, this 
restriction is appropriate for large graphs where it is difficult to account for many recovering functions, and it is also well suited for the analysis of storage capacity because it enables us to focus on shift invariant systems, and paves the way for using insights from symbolic dynamics, which we mention below.

Let $Q$ be a finite alphabet and denote by $Q^*$ the set of all finite words over $Q$. 
We say that a finite word $w$ over $Q$ is a \textbf{subword} in $x\in\Q^{\Z}$, 
denoted $w\prec x$, if $w$ appears anywhere in $x$ as a sequence of consecutive letters.

\begin{definition}\label{def:cap} Let $X$ be a $(k,l)$-recoverable system over a finite alphabet $Q$ of size $|Q|=q$. Let $\cB_n(X)$ 
denote the set of all length-$n$ words each of which appears as a subword in some $x\in X$. The \textbf{capacity} of $X$ is defined as
   $$
   \ccap(X)=\lim_{n\to\infty} \frac{1}{n}\log_q \abs{\cB_n(X)}
   $$
   where the limit exists due to the sub-additivity of $\log_q \abs{\cB_n(X)}$ and Fekete's lemma\footnote{Fekete's lemma states that a sequence $\{a_n\}$ such that $a_{m+n}\le a_m +a_n$ necessarily has a limit, and $\lim\frac{a_n}n=\inf_n\frac{a_n}n$ (see \cite{Fek1923,BruErd1952}).}.
\end{definition}

Recoverable systems are strongly related to the well-established notion of {\em constrained systems} \cite{MarRotSie98,LinMar21}. Informally speaking, a constrained system is a set of all sequences over a finite alphabet that do not include subwords from a set $\cF$ of forbidden words. As explained below, a $(k,l)$-recoverable system is a sub-system of a constrained system. Indeed, for given $k,l\in \mathbb{N}$ the $(k,l)$-recoverability property gives rise to a set of forbidden words $\cF$ (by ruling out conflicts in recovery), and bi-infinite sequences without forbidden subwords form a subset in a constrained system.

The motivations to study $(k,l)$-recoverable systems are multifold. Indeed, they form  a natural extension of the notion of storage codes to the infinite setting, and enable one to bring in methods from constrained systems such as de Bruijn graphs and Markov chains, which have not so far been used for storage codes. Recoverable systems also yield new insights into constructions of storage 
(index) codes for some classes of finite graphs, notably, circulant graphs. For such graphs we suggest a method of constructing a storage code over any given alphabet, yielding a lower bound on the maximum achievable code rate. We also show that capacity of recoverable systems (and related storage codes) can be increased by defining a probabilistic relaxation of the above definitions, which is a new concept in the area of erasure codes on graphs. This concept brings forth a link between recoverable systems and entropy of dynamical systems, enabling one to use tools 
such as metric entropy to derive bounds on the system capacity. It also suggests a group of new problems with a potential for connecting a broad array of methods from ergodic theory and discrete dynamical systems with
the context of storing and recovering information on large graphs.

Our results are presented in the next three sections. We begin in Sec.~\ref{sec:general} with establishing general properties related to the capacity of recoverable systems. In Sec.~\ref{sec:constructions} we present two constructions of $(1,1)$-recoverable systems, one of which attains the maximum possible capacity whenever $q$ is a whole square. Both constructions use techniques from constrained systems, namely the description 
of a system in terms of a Markov chain and de Bruijn graphs.

In Sec.~\ref{sec:relax} we study a relaxation of the $(k,l)$-recoverability property wherein the recovered version of the contents is permitted to have a small residual entropy, resulting in a probabilistic version of the contents recovery. 
We call systems arising in this way {\em $(\epsilon,k,l)$-recoverable}. In this part it is natural to switch to measure-theoretic language, making each vertex to carry a random variable $X_i$ such that the collection of random variables $(X_i,i\in\Z)$ follows some joint distribution. We provide a characterization of measures that maximize the entropy of such systems. We also show how to obtain a measure supported on an $(\epsilon,k,l)$-recoverable system from a $(k,l)$-recoverable system with 
maximum capacity. The assumption of a single recovery function for every vertex on the line translates into a characterization of capacity  as the topological entropy of the system. This allows us to use the variational principle which relates topological entropy and metric entropy, thus enabling us to borrow tools from entropy theory to evaluate the capacity. 

Our results in this part are as follows. We prove that there exists an entropy-maximizing measure $\mu$, which turns out to be a Markov measure whose conditional Shannon entropy equals $\epsilon.$ 
We propose a procedure of constructing an $\epsilon$-recoverable measure from a deterministic recoverable
system and bound below the increase of the system capacity due to the relaxation. 

In the concluding Section \ref{sec:conclusion} we present some details regarding the connection between $(k,l)$-recoverable systems and storage codes, and mention several problems for future research.

\section{General properties of recoverable systems}\label{sec:general}
We establish properties of $(k,l)$-recoverable systems by linking them to constrained systems, beginning with some notation. 
Similarly to the infinite case, if $w,u\in Q^*$ are finite words such that $w$ appears in $u$ as
a consecutive subsequence, we say that $w$ is a subword of $u$ and denote this $w\prec u$. Let $|w|$ be the length of the word $w$.
The concatenation of finite words $w$ and $u$ is denoted by $wu$ and in particular, $w^n, n\ge 2$ is an $n$-fold concatenation of $w$ with itself. Let $T$ be a shift of the infinite word $x\in Q^\Z$ to the left, so for every $i\ge 1, j\in \Z$ and any $x\in  Q^{\Z}$ we have $(T^i x)_j=x_{i+j}$. 

The notion of recoverable systems is closely related to constrained systems. In what follows we will present some relevant notation and results from constrained systems (for more details see \cite{LinMar21,MarRotSie98})

Let $G=(V,E,L)$ be a directed graph, where $V$ is a finite set of vertices, $E$ is the set of directed edges, and $L:E\to Q$ is a label 
function that assigns symbols from $Q$ to edges, one symbol per edge. 
A path $\gamma$ in $G$ is a finite sequence of edges $\gamma=(e_0,\dots,e_{n-1}).$ The label of the path $L(\gamma)$ is the sequence of symbols from $ Q$ obtained by reading off the labels of the edges in $\gamma$, i.e., $L(\gamma)=(L(e_i))_{i\in [n]}$. We allow multiple edges between a pair of vertices, but require that they carry different labels.

A \textbf{constrained system} $\cS$ is the set of all finite words given by the labels of the paths in a directed labeled graph $G$. 
We call the graph $G$ a {\em presentation} of $\cS,$ and we say that $\cS$ is presented by $G$. 

It is well known that a constrained system can also be defined by a (possibly infinite) set $\cF\subseteq  Q^*$ of finite words, called {\em forbidden words}, in the following way. Let $X_{\cF}\subseteq Q^{\Z}$ be the set of all bi-infinite sequences such that no $x\in X_{\cF}$ contains any word from $\cF$ as a subword, i.e., there are no $i,j\in\Z$ for which $(T^i x)_{[j]}\in\cF$. 
For every constrained system $\cS$, it is possible to find a corresponding $\cF$ such that $\cS=\bigcup_{n\in \N}\cB_n(X_{\cF})$ (see \cite{LinMar21}), where $\cB_n(X_{\cF})$ is the set of all length-$n$ words that appear as a subword in some $x\in X_{\cF}$. Therefore, both $\cS$ or $X_{\cF}$ will be referred to as a constrained system. Note however that they are objects of different kinds: 
while $\cS$ is a set of finite words, $X_{\cF}$ is a set of bi-infinite sequences.

The graph $G=(V,E,L)$ that presents a constrained system $X_{\cF}$ can be described explicitly in the case that all the forbidden words 
have the same length, i.e., $\cF\subseteq Q^n$ for some $n\in \N.$ 
Namely, take $V=Q^{n-1}$ to be the set of all words of length $n-1$ over $Q.$ Draw an edge from 
$u=(u_0,\dots,u_{n-2})$ to $v=(v_0,\dots,v_{n-2})$ if $v_{i}=u_{i+1},\; i\in [n-2]$ and $u_0 \dots u_{n-2} v_{n-2}\notin \cF$. The edge $(u,v)$ carries the label $L((u,v))=v_{n-2}$. The sequences $x\in X_{\cF}$ correspond to the sets of labels of the paths in $G$.

The capacity of a constrained system $\cS$ is defined similarly to the capacity of a $(k,l)$-recoverable system.
Denote by $\cB_n(\cS)$ the set of all length-$n$ words in $\cS$, i.e., $\cB_n(\cS):= \cS \cap  Q^n$. 
Similarly, if the system is given by $X_{\cF}$ then $\cB_n(X_{\cF})$ denotes the set of all length-$n$ words that appear as subwords in some $x\in X_{\cF}$. 
The \textbf{capacity} of a constrained system $\cS$ is defined as \cite{LinMar21}
          $$
\ccap(\cS)=\lim_{n\to\infty} \frac{1}{n} \log_{q} |\cB_n(\cS)|.
         $$
In the language of symbolic dynamics this quantity is also known as the {\em topological entropy} of the system; see \cite[Sec.~3.1]{KatokHasselblat1995} or \cite[Ch.~7]{Walters1982}. We remark that we use the same notation for the capacity of constrained systems and recoverable systems. This usage is justified by Lemma~\ref{lem:rec_sys_subset_con} below.

It is well known that the limit in the definition of $\ccap(\cS)$ exists due to sub-additivity of the quantity $\log_q |\cB_n(\cS)|$ and Fekete's lemma. Moreover, 
	\begin{equation}\label{eq:inf}
	\ccap(\cS)=\inf_{n\in\N} \frac{1}{n} \log_{ q} |\cB_n({\cS})|.
	\end{equation}

Assume that a constrained system $\cS$ is presented by a graph $G=(V,E,L).$ The adjacency matrix $A_G$ 
is a $|V|\times |V|$ matrix such that $(A_G)_{(u,v)}$ is the number of edges from $u$ to $v$ in $G$. 
If the graph $G$ is {\em strongly connected} (there is a directed path between any two vertices), then the capacity $\ccap(\cS)$ can be calculated 
using $A_G$. Indeed, the Perron-Frobenius theorem states that the largest eigenvalue of $A_G$, denoted $\lambda,$ is real, simple, and positive with strictly positive left and right eigenvectors. Then the capacity of $\cS$ equals $\ccap(\cS)=\log_{q} \lambda$ \cite[Theorem 3.9]
{MarRotSie98}. In fact, $\ccap(\cS)=\log_{q} \lambda$ even if the graph $G$ is not strongly connected (see \cite[Theorem 4.4.4]{LinMar21})

To analyze recoverable systems, we relate them to constrained systems using the following simple idea. Notice that if 
a word $w\in Q^k$ can be recovered from its $l$-neighborhood $u,v\in Q^l$, then no sequence $x\in X$ can contain a subword $uw'v$
for any $w'\in Q^k, w'\ne w.$ This means that the words $uw'v$ are forbidden as subwords in some constrained system. 
This observation suggests the following definition.
\begin{definition}\label{def:confusable}
	Let $\cF\subseteq  Q^*$ be a set of finite words and let $k,l\in\N$. We say that $\cF$ is a $(k,l)$-\textbf{admissible set} if for any $u,v\in Q^l$, 
	$$
	\abs{\mathset{w\in Q^k ~:~ uwv\prec s\in \cF }}\geq q^k-1.
	$$
\end{definition}

It is immediate that a constrained system $X_{\cF},$ where $\cF$ is a $(k,l)$-admissible set, is a $(k,l)$-recoverable system. 
\begin{lemma}
	\label{lem:rec_sys_subset_con}
	Any $(k,l)$-recoverable system $X\subseteq  Q^{\Z}$ is a subset of a constrained system $X_{\cF}$ for some $(k,l)$-admissible set $\cF$. Conversely, any constrained system $X_{\cF}$ with $\cF$ a $(k,l)$-admissible set is a $(k,l)$-recoverable system.
\end{lemma}

\begin{IEEEproof}
	Let $X$ be a $(k,l)$-recoverable system over the alphabet $ Q$. Now let $\cF= Q^{2l+k}\setminus \cB_{2l+k}(X),$ then $\cF$ is a $(k,l)$-admissible set and that $X\subseteq X_{\cF}$. 
Indeed, if not, then there are two distinct words $w_1, w_2 \in Q^k$
such that $uw_1v\prec x$ and $uw_2v\prec y$, and by shift invariance we may assume that $x_{[k]}=w_1$ and $y_{[k]}=w_2$. This contradicts
the recovery condition. Finally, the converse statement is obvious by Def.~\ref{def:confusable}.
\end{IEEEproof}

Let $Y$ be a $(k,l)$-recoverable system over an alphabet $Q$ of size $q$. As just argued, $Y$ is a subset of some constrained system $X_{\cF},$ where $\cF$ is a $(k,l)$-admissible set, and thus the quantity
    \begin{equation}\label{eq:max}
    C_q(k,l):= \max_{\substack{Y\subseteq  Q^{\Z} \\ Y \text{ is } (k,l)\text{-recoverable}}} \ccap(Y).
    \end{equation}
is well defined.  
Moreover, $\ccap(Y)\leq \ccap(X_{\cF}),$ and since there is a finite number of possible $(k,l)$-admissible sets over $ Q$, 
the maximum in \eqref{eq:max} is attained.  We say that $X$ has \textbf{maximum capacity} if $\ccap(X)=C_q(k,l)$ and note that the maximizing system is 
not necessarily unique.

The following properties of $C_q(k,l)$ are immediate: if $l_1\ge l_2$ or $q_1\ge q_2,$  then
    \begin{equation}\label{eq:monotone}
      C_{q_1}(k,l_1)\ge (\log_{q_1}q_2)C_{q_2}(k,l_2)
    \end{equation}
To illustrate these definitions, let us consider the simplest instance of the concepts introduced above, namely binary $(1,1)$-recoverable systems.
\begin{proposition}\label{ex:1}
The capacity $C_2(1,1)\approx0.4057,$ and there is exactly one system $X$ that attains it, up to symbol relabeling.
\end{proposition}

\begin{IEEEproof}	
The proof proceeds by examining all the possible binary $(1,1)$-recoverable systems. On account of Lemma 
\ref{lem:rec_sys_subset_con} this amounts to examining all the possible constrained systems $X_{\cF},$ where
$\cF$ is a $(1,1)$-admissible set. 
The $(1,1)$-recoverability condition implies that a symbol is determined completely by the value of its 
neighbors. Thus, for instance, one of the sequences $000, 010$ is in $\cF$. Overall, out of the eight possible
triplets four are in $\cF$, while the remaining four appear in the sequences in $X.$ 

If $010$ or $101$ are in $\cF,$ then $\ccap(X)=0$. This is because in either of these cases, $X$ is formed of at most 
two sequences. Indeed, suppose that $010\in\cF$. Between the two triplets $111$ and $101$, one must be forbidden to allow
recoverability. Say it is $111$, then $X=\{(000)^{\Z},(101)^{\Z}\}$ (the word $(0011)^{\Z}$ is forbidden to allow recoverability). Alternatively, if $101$ is forbidden,
then $X=\{(000)^{\Z},(111)^\Z\}.$  The same arguments apply to $101$ upon the relabeling $0\to1, 1\to0.$
Thus $010,101\notin\cF,$ which implies that $000,111\in\cF$.

It remains to analyze the triplets 110 and 100. If $110\in\cF$, then by shift invariance also $011\in\cF$, otherwise the capacity is $0$. 
This yields that $000,111,110,011\in\cF$ and the remaining four patterns are allowed. 
This system is known as a $(1,2)$-RLL constrained system, and its capacity is approximately $0.4057$ \cite[Ch. 4]{Imm2004}. The pattern 100 yields the same result by relabeling. This concludes the proof.
\end{IEEEproof}

The following result enables us to claim that $C_q(k,l)$ is positive for general alphabets.

\begin{corollary}
	\label{cor:pos_ent}
	Let $ q\geq 3$ and let $2\le k\le l-1$, then
	 $$
	 C_q(k,l)\ge  C_q(k,k+1)\ge \frac{\log_q 2}{k+2}.
	 $$
\end{corollary}

\begin{IEEEproof} 
    We explicitly describe a $(k,k+1)$-recoverable system $X$ over $Q=\mathset{0,1,2}$ with positive capacity. By 
\eqref{eq:monotone} this will imply the general claim.
  
Construct a set of sequences $X$ by taking all bi-infinite sequences over $\mathset{a,b}$ and replacing each $a$ with  $2\, 1^{k+1}$ 
(two followed by $k+1$ ones) and each $b$ with $2\, 0^{k+1}$. We claim that $X$ is a $(k,k+1)$-recoverable system. Indeed, let 
$u=(u_0\dots u_k),v=(v_0\dots v_k)\in Q^{k+1}$, $w\in Q^k$ and let us show that $w$ is recoverable from $u$ and $v$. 
Let $u_{-1}$ be the symbol right before $u_0.$ Exactly one of the elements $u_{-1},u_0,\dots,u_k$ is 2, and this gives rise to
the following 4 cases.
\begin{enumerate}
\item[$(i)$] $u_{-1}=2,$ then $w=2v_0^{k-1};$
\item[$(ii)$] $u_0=2,$ then $w=u_{k}2v_0^{k-2};$
\item[$(iii)$] $u_i=2, 1\le i\le k-1,$ then $w=u_k^{i+1}2v_0^{k-i-2};$
\item[$(iv)$] $u_k=2,$ then $w=v_0^k.$
\end{enumerate}
This shows that $w$ can be recovered from its $(k+1)$-neighborhood. 
    
To calculate $\ccap(X)$, notice that  $2^n=|\mathset{a,b}^n|=|\cB_{n(k+2)}(X)|$. Thus, 
    $$
    \frac{1}{n(k+2)}\log_q |\cB_{n(k+2)}(X)|=\frac{\log_q 2}{k+2}.
    $$
Since this equality is true for any $n$, this implies the claimed result.
\end{IEEEproof}
    
Our next result is an upper bound on the capacity of $(k,l)$-recoverable systems. 
\begin{lemma}
	\label{lem:upper_bound_ent}
For all $q\ge 2$ and $k\le l$ we have
	$$
	C_q(k,l) \leq \frac{l}{k+l}.
	$$
\end{lemma}

\begin{IEEEproof}
	From Lemma \ref{lem:rec_sys_subset_con} it suffices to show that $\ccap(X_{\cF})\leq \frac{l}{k+l}$ for every constrained system $X_{\cF}$ with a $(k,l)$-admissible set $\cF$. On account of \eqref{eq:inf} we need to prove that for every $\epsilon>0$, there exists $n\in\N$ for which 
	$$
	\frac{1}{n} \log_{ q} |\cB_n(X_{\cF})|\leq (1+\epsilon)\frac{l}{k+l}.
	$$
	
	Fix $\epsilon>0$ and take $m\in\N$ large enough such that 
	$\frac{m+1}{m} < 1+\epsilon$. Now take $n=m(k+l)+l$ and let $F_n$ denote the set of indices given by
	$$
F_n=\mathset{j\in [n] \,|\, j\bmod (k+l) \in [l]}.
	$$
Consider $x\in X_{\cF}$. If the symbols $x_j, j\in F_n$ are known, then the remaining symbols in
$[n]\backslash F_n$ are determined uniquely because of the recoverability property.
Since there are at most $ q^{(m+1)l}$ different words that can appear in the locations $F_n$, we obtain 
	\begin{align*}
	\frac{1}{n} \log_{ q} \parenv{|\cB_n(X_{\cF})|}&\leq \frac{1}{n}\log_{ q}  q^{(m+1)l}\\
	&\le \frac{(m+1)l}{m(k+l)} \\
	&\le  (1+\epsilon)\frac{l}{k+l}.
	\end{align*}
\end{IEEEproof}
An obvious observation, implied by this bound, is that many symbols cannot be recovered from a small-size
neighborhood, i.e., $\lim_{k\to\infty} C_q(k,l)=0.$

For certain values of $q$ we can use the well-known edge covering construction from the storage coding and index coding literature \cite{MazMcgVor2019,BKL2013} to attain the bound in this lemma. 
\begin{proposition}
\label{prop:block_enc_prop}
If $q$ is a whole square, then
    $C_q(l,l)=\frac{1}{2}.$ If $q=t^{k+1}$ for some integer $t,$ then $ C_q(k,1)=\frac{1}{k+1}.$
\end{proposition}
\begin{IEEEproof} 
Consider a graph $G(V,E)$ with $V=\Z$ and $(i,j)\in E$ if and only if $|i-j|=1.$
Place a symbol of the alphabet $Q$ on each edge and assign to a vertex the pair of symbols which appear on the edges adjacent to it,
then the symbol $x_i, i\in\Z$ is determined once we know $x_{i-1}$ and $x_{i+1}.$ This gives rise to a $(1,1)$-recoverable system
$X$ over the alphabet of size $q^2,$ where $q=|Q|,$ and clearly $|\cB_n(X)|\ge q^n.$ If $q=t^2,$ then
we conclude that 
   $$
   C_q(1,1)\ge \lim_{n\to\infty}\frac 1n \log_q t^n= 1/2.
   $$
Lemma \ref{lem:upper_bound_ent} implies that this is an equality. 

For $l\ge 2,$ assign the vertex $x_i,\; i\in \Z$ the pair of 
symbols that appear on the edges $(i-1,i)$ and $(i+l-1,i+l). $

As before, this enables us to argue that $C_q(l,l)=\frac 12.$ To prove the second claim of the lemma, construct a system $X$ by assigning each vertex $i\in \Z$ the $k+1$-tuple that appears on the edges 
   $$
   (i-1,i),(i,i+1),\dots,(i+k-1,i+k).
   $$
It is straightforward to check that the system $X$ is $(k,1)$-recoverable and that its capacity is at least $1/(k+1).$
Now the claim follows from Lemma \ref{lem:upper_bound_ent}.
\end{IEEEproof}

We can remove the assumptions on $q$ in the limit of large alphabet.
\begin{corollary} \label{th:lim_ent_exists}
For $l,k\ge 1$ and $q\to\infty$ the limits of $C_q(l,l)$ and $C_q(k,1)$ exist, and
	\begin{align*}
		\lim_{q\to\infty} C_q(l,l)&=\frac{1}{2} \\
		\lim_{q\to\infty} C_q(k,1)&=\frac{1}{k+1}.
	\end{align*}
	\end{corollary}
\begin{IEEEproof} 
Letting $q=t^2+r,$ we find
  $$
  C_q(l,l)\geq \lim_{n\to\infty} \frac{1}{n} \log_{t^2 +r} t^n= \log_{t^2 +r} t,
  $$
  which goes to $1/2$ as $q\to\infty$. 
Similarly, letting $q=t^{k+1}+r,$ we find
  \begin{align*}
  C_q(k,1)&\geq \lim_{n\to\infty} \frac{1}{n} \log_{t^2 +r} t^n \\
  &= \log_{t^{k+1} +r} t\to \frac{1}{k+1}.
  \end{align*}
Now the claim follows from Lemma \ref{lem:upper_bound_ent}.
\end{IEEEproof}

%

\section{Constructions of $(1,1)$-recoverable systems}\label{sec:constructions}
In this section we focus on the case $k=l=1.$ For this case, the capacity is at most $1/2$ by Lemma \ref{lem:upper_bound_ent}, and we have pointed out that it is attainable when $q$ is a whole square. 
By extending the edge covering construction to all $q$ and using \eqref{eq:monotone} we can obtain lower bounds on 
$C_q(1,1)$ for general $q$. 
In this section we present two different constructions of recoverable systems that yield higher capacity values for all $q$ except whole squares.

\subsection{A recursive construction}\label{sec:rec}
In the first construction we relate $C_q(1,1)$ to $C_{q-2}(1,1).$ 
Let $G=(V,E)$ be a strongly connected graph and  let $\Gamma_n(G)$ be the set of all directed paths of length 
$n\ge 1$ in $G$. Let $p$ be a probability vector $p=(p_0,\dots,p_{|V|-1})$ and let $\epsilon>0$. We denote by 
$\Gamma_n^{\epsilon}(G,p)$ the set of all directed paths $\gamma\in\Gamma_n(G)$ such that for every $v\in V$, $\gamma$ traverses $v$ at least $(p_v-\epsilon)n$ times. 
The following result is cited from \cite{MarRotSie98}; we have adjusted it here to our notation.

\begin{lemma}[{\rm{\!\!\cite[Thm. 3.15]{MarRotSie98}}}]
	\label{lem:stat_prob1}
	Let $G=(V,E)$ be a strongly connected graph. There exists a probability vector $p=(p_0,\dots,p_{|V|-1})$ such that for every $\epsilon>0$, there exists an $n_{\epsilon}\in\N$ such that for all $n\geq n_{\epsilon}$, 
	$$
	\abs{\Gamma_n^{\epsilon}(G,p)}\geq (1-\epsilon)|\Gamma_n(G)|.
	$$
\end{lemma}
 
 The idea of the construction, formalized in the next lemma, consists in adding two new symbols to the
 alphabet and adding a loop of length four to a frequently traversed vertex in the graph $G$ that presents 
 the system.
\begin{lemma}
	\label{lem:lower_bound_m}
	We have
	\begin{align*}
	C_{q+2}(1,1)\geq 
	C_{q}(1,1) \log_{q+2} q +\frac{1}{q^2}\log_{q+2} \parenv{1+\frac{1}{q^2}}.
	\end{align*}
\end{lemma}
\begin{IEEEproof} Let $X$ be a $(1,1)$-recoverable system with maximum capacity over $Q$.
By Lemma \ref{lem:rec_sys_subset_con} we may assume that $X$ is a constrained system $X_{\cF}$ for some $(1,1)$-admissible set $\cF\subseteq Q^3$ of forbidden words. 
Let $G_q=(V_q, E_q, L_q)$ be a graph that presents $X_{\cF}$ and suppose that each vertex in 
$G_q$ corresponds to a pair of symbols of $Q$. 
For two vertices $u=(u_1u_2), v=(v_1v_2)$, the edge $(u,v)\in E_q$ iff $u_2=v_1$ and 
$u_1u_2v_2\notin \cF$. The label of the edge $L_q\parenv{(u,v)}=v_2$. We may also assume that $G_q$ is connected 
since the capacity can be calculated by considering only the connected components of $G_q$. 
	
	Fix $\epsilon>0$ and use Lemma \ref{lem:stat_prob1} to find $n_{\epsilon}\in\N$ and a probability vector $p=(p_0,\dots,p_{|V_q|-1})$ such that for $n\geq n_{\epsilon}$
	\[
	\abs{\Gamma_n^{\epsilon}(G_q,p)}\geq (1-\epsilon)|\Gamma_n(G_q)|.
	\]
	Since $p$ is a probability vector, there exists a vertex $v\in V_q$ for which $p_v\geq \frac{1}{q^2}$. Assume that $v$ corresponds to the pair $(ab)\in Q^2$.
	
Let us add symbols $\alpha$ and $\beta$ to the alphabet $Q.$ Construct a new graph $G_{q+2}=(V_{q+2},E_{q+2},L_{q+2})$
by adding the vertices $u_1=(b\alpha), u_2=(\alpha\beta), u_3=(\beta a)$ with edges 
	\begin{align*}
	&\xi_1=(ab,u_1),\xi_2=(u_1,u_2),\\
	&\xi_3=(u_2,u_3), \xi_4=(u_3,ab)
	\end{align*}
	and labels $\alpha,\beta,a,b$, respectively. The graph $G_{q+2}$ presents a system $X_{\cF'}$ for some $\cF'$ over the alphabet of size $q+2.$
	
Let us show that $X_{\cF'}$ is a $(1,1)$-recoverable system. Consider a sequence $x\in X_{\cF'}$. 
Note that if $x\notin X_\cF,$ it must contain at least one of the symbols $\alpha,\beta.$
 By the construction of $G_{q+2}$, 
the words of length three that contain $\alpha$ or $\beta$ and can appear as subwords in $x,$ are 
	\[
	A:=\mathset{ab\alpha, b\alpha\beta,\alpha\beta a, \beta ab}.
	\]
	Therefore, the system $X_{\cF'}$ is a constrained system that corresponds to the set of forbidden words  
	\[
	\cF'=\parenv{(Q\cup\{\alpha,\beta\})^3\setminus (Q^3\cup A)}\cup \cF.
	\]
	It is straightforward to check that $\cF'$ is a $(1,1)$-admissible set, which implies that $X_{\cF'}$ is a $(1,1)$-recoverable system.
	
Let us bound below the capacity of $X_{\cF'}$. 
Assume first that $p_v<\frac{1}{4}$ and let $r\in\N$ be large enough such that $\floorenv{(1-4p_v)r}\geq 
n_{\epsilon}$. Note that if $\gamma=(e_0,\dots, e_r)$ is a path in $G_q$, then it is also a path in $G_{q+2}$. Moreover, if for some $i\in [r]$, $e_i$ terminates in $v$, then  
	\[
	\gamma'=(e_0,\dots, e_i,\xi_1,\xi_2,\xi_3,\xi_4,e_{i+1},\dots, e_r)
	\]
	is a path in $G_{q+2}$ of length $|\gamma|+4$. Let $\cB_r^{\epsilon}(X_{\cF})$ denote the set of words in $\cB_r(X_{\cF})$ with at least $M_{v}:=\lfloor(p_v-\epsilon)r\rfloor$ appearances of $ab$. Each such word is obtained by reading off the labels of a length-$n$ path in $G_q$. Then
	\begin{align*}
	|\cB_r(X_{\cF'})|&\geq 	\sum_{i=0}^{M_{v}} \binom{M_{v}}{i}  \abs{\cB_{r-4i}^{\epsilon}(X_{\cF})} \\
	&\stackrel{(a)}{\geq} \sum_{i=0}^{M_{v}} \binom{M_{v}}{i}  (1-\epsilon)q^{(r-4i)\ccap(X_{\cF})}\\
	&= q^{r \ccap(X_{\cF})} (1-\epsilon) \sum_{i=0}^{M_{v}} \binom{M_{v}}{i} q^{-4i\ccap(X_{\cF})} \\
	&= q^{r \ccap(X_{\cF})} (1-\epsilon) \parenv{1+q^{-4\ccap(X_{\cF})}}^{M_{v}}
	\end{align*}
	where to claim inequality $(a)$ we used the estimate $|\cB_{r-4i}^{\epsilon}(X_{\cF})|\geq (1-\epsilon)|\cB_{r-4i}(X_{\cF})|$, valid for all $i\leq (p_v-\epsilon)r$, the assumption on $r$, and definition  \eqref{eq:inf}.
We obtain 
	\begin{multline*}
	\frac{1}{r} \log_{q+2} |\cB_r(X_{\cF'})| \geq \ccap(X_{\cF}) \log_{q+2} q \\
	+ \frac{(p_v-\epsilon)r-1} {r}\log_{q+2} (1+q^{-4\ccap(X_{\cF})}) \\
	+\frac{1}{r} \log_{q+2} (1-\epsilon).
	\end{multline*}
	From Lemma \ref{lem:upper_bound_ent}, $\ccap(X_{\cF})\leq \frac{1}{2}$. Since $p_v\geq \frac{1}{q^2}$, taking $r\to\infty$ we obtain 
	\begin{align*}
	\ccap(X_{\cF'})\geq 
	\ccap(X_{\cF})\log_{q+2} q+\parenv{\frac{1}{q^2}-\epsilon}\log_{q+2} \parenv{1+\frac{1}{q^2}}.
	\end{align*}
	Since $\epsilon$ is arbitrary, 
	\begin{align*}
	    \ccap(X_{\cF'})\geq 
	    ccap(X_{\cF}) \log_{q+2} q+\frac{1}{q^2}\log_{q+2} \parenv{1+\frac{1}{q^2}}.
	\end{align*}
	
	The case $p_v\geq \frac{1}{4}$ is similar. We only consider the first $n/4$ visits to $v$, with $r$ large enough such that $\floorenv{\epsilon r}\geq n_{\epsilon}$, $M_v:=\floorenv{(\frac{1}{4}-\epsilon)r}$, and otherwise the proof is the same. 
\end{IEEEproof}

For alphabets of size $q$ not a whole square, i.e., $q= t^2+r$ with $r>0$, we have two options of constructing
a recoverable system, namely relying on the alphabet of size $t^2$ (by the edge covering procedure, see the proof of 
Prop.~\ref{th:lim_ent_exists}), 
or using the construction given in the
proof of Lemma \ref{lem:lower_bound_m}. The following example shows that the latter approach can be
superior to the former, yielding a system with larger capacity. 

\begin{example}
{\rm    Let $X$ be a $(1,1)$-recoverable system over an alphabet of size $q=6$, constructed by edge covering. 
 This yields the estimate $\ccap(X)=\log_6 2$. At the same time, 
 from Lemma \ref{lem:lower_bound_m} we obtain
	$$
	C_6(1,1)\geq \frac{1}{2} \log_{6} 4 +\frac{1}{16}\log_{6}\frac{17}{16}> \log_6 2.
	$$}
\end{example}

\subsection{Construction via de Bruijn graphs}
We now provide a second construction of $(1,1)$-recoverable systems over an alphabet $Q$ of size $q$, which relies on the de Bruijn graphs \cite{DeB1946,Gol67}. The capacity analysis of the constructed systems will be carried out for the case 
of $q$ close to the nearest whole square greater than it.
 For a finite alphabet $Q$ and $d\in\N$, a \textbf{de Bruijn graph} of order $d$ over $Q$ is the directed 
 labelled 
 graph $\db( Q,d)=(V,E,L),$ whose vertex set $V=Q^d$ corresponds to length-$d$ words over $Q$. 
 An edge $(u,v)\in E$ if $v=(v_0\dots v_{d-1}), u=(u_0\dots u_{d-1})$ and $v_i=u_{i+1}$ for all $i\in [d-1]$, and the label of $(u,v)\in E$ is $L((u,v))=v_{d-1}$. In words, there is an edge from $u$ to $v$ if $v$ is 
 the $(d-1)$-tail of $u$ appended by some symbol $v_{d-1}\in Q$ which serves as a label.
 The relevance of the de Bruijn graphs to $(k,l)$-recoverable systems is established in the next lemma.
\begin{lemma}
	\label{lem:db_d_recov}
	For a finite alphabet $ Q$, $\db( Q,d)$ yields a presentation of a $(k,1)$-recoverable system over $Q^d$ for every $k\in [d]$.
\end{lemma}

\begin{IEEEproof}
	Consider a new labeling function $L'$ such that for an edge $(u,v)\in\db(Q,d)$, 
	$L'((u,v))=(u_0,\dots,u_{d-1})\in Q^d$. By definition, $\db( Q,d)$ with the labeling $L'$ is a presentation of a constrained system $\cS$ over $Q^d$.
	
	To see that the system $\cS$ is $(k,1)$-recoverable for every $k\in [d]$, notice that if $x$ is a sequence obtained by reading off the labels of a path in $\db(Q,d)$ (with $L'$), then $\bfx_{-1}=(x_{-1},x_0,\dots,x_{d-2})$ and $\bfx_k=(x_k,x_{k+1},\dots,x_{k+d-1})$. While $k\leq d-1$ we have that for every $i\in [k]$, the sequences
$\bfx_i=(x_i,\dots,x_{i+d-1})$  can all be constructed from $\bfx_{-1}$ and $\bfx_k$, proving recoverability.
\end{IEEEproof}

\begin{remark}
  The recovery procedure used in the proof relies on sharing the label between a pair of vertices across the edge that connects them, and is similar to the edge covering construction used in the proof of Proposition \ref{prop:block_enc_prop}. In fact, a de Bruijn graph $\db(Q,d)$ can be though of as a presentation of a 
  "$d$-block-encoder". In terms of Proposition \ref{prop:block_enc_prop}, if $G(V,E)$ is a graph with $V=\Z$, $(i,j)\in E$ if $|i-j|=1$, and a symbol is placed on each edge, the $d$-block-encoding is the system obtained by assigning each vertex $i\in \Z$ the $d$-tuple that appears on the edges 
  \[(i-d,i-d+1),\dots,(i-2,i-1),(i-1,i).\]
  Thus, the capacity of the $(k,1)$-recoverable system presented by $\db(Q,d)$ is $\frac{1}{d}$.
\end{remark} 

To conform with the definition of a de Bruijn graph, we fix the number of vertices to $q^d.$ At the same time, any subgraph of $\db( Q,d)$ corresponds to a $(k,1)$-recoverable systems with fewer than $q^d$ vertices.  We state this observation in the next corollary.
  \begin{corollary}	\label{cor:sub_db_recov}
	Any subgraph of $\db( Q,d)$ presents a $(k,1)$-recoverable system for any $k\in [d]$. 
  \end{corollary}
By relabeling the vertices, we can obtain a $(k,1)$-recoverable systems over an alphabet of size 
strictly smaller than $q^d$. This is the key insight of our construction. 

The construction makes use of de Bruijn graphs of order $d=2$. Therefore, unless specified otherwise,
for the rest of the section we assume that $d=2$ and write $\db(Q)$ instead of $\db(Q,2)$.
We will need some properties of the adjacency matrix $A_D$ of $\db(Q),$ assuming that the vertices are ordered
according to the lexicographic order of the their labels. The following description of 
$A_D$ is immediate.
\begin{lemma} \label{lem:shape_db_adj_mat}
Let $A_D$ be the adjacency matrix of $\db( Q)$. Then for $i,j\in \sparenv{q^2}$    
	\[(A_D)_{(i,j)}=\begin{cases}
	1 \text{ if }  j\in \{(i\bmod q)\cdot q +[q]\}\\
	0 \text{ otherwise}.
	\end{cases}.
	\]
\end{lemma}

\begin{IEEEproof} Interpret the words of length $2$ over $Q$ as base-$q$ numbers from $0$ to $q^2-1.$
Clearly, a number $u$ is connected to a number $v$ if and only if $v\in \{(qu \bmod q^2) +[q]\}.$
\end{IEEEproof}

\begin{example}{\rm The matrices $A_D$ have a simple structure which we show in examples. This makes later proofs
easier to follow.
Let $A_{D_i},i=1,2$ be the adjacency matrices of $D_1=\db([2])$ and $D_2=\db([3]),$ then
	\begin{align*}
	&A_{D_1}=\begin{bmatrix}
	1&1&0&0 \\
	0&0&1&1 \\
	1&1&0&0 \\
	0&0&1&1
	\end{bmatrix} \\
	&\text{ and }\\ 
	&A_{D_3}=\begin{bmatrix}
	1&1&1&0&0&0&0&0&0 \\
	0&0&0&1&1&1&0&0&0 \\
	0&0&0&0&0&0&1&1&1 \\
	1&1&1&0&0&0&0&0&0 \\
	0&0&0&1&1&1&0&0&0 \\
	0&0&0&0&0&0&1&1&1 \\
	1&1&1&0&0&0&0&0&0 \\
	0&0&0&1&1&1&0&0&0 \\
	0&0&0&0&0&0&1&1&1
	\end{bmatrix}.
	\end{align*}}
\end{example}

The following property of de Bruijn graphs is well known (e.g., \cite[Exercise 1.5]{BH2012}). We include a short
proof for completeness.
\begin{lemma}
	\label{lem:eig_sys_db_graph}
	Let $A_D$ be the adjacency matrix of $\db(Q)$. Then the eigenvalues of $A_D$ are 
	$\lambda_0=0$ with multiplicity $q^2 -1$ and $\lambda_1=q$ with multiplicity 1.
\end{lemma}
\begin{IEEEproof} 
	In $\db(Q)$ there is a unique path of length $2$ from any $v=(v_0,v_1)\in V$ to any $u=(u_0,u_1)\in V$, namely, $\gamma=(v,w_1),(w_1,u)$ where $w_1=(v_1u_0)$.
	Thus, $(A_D)^2=J,$ the all ones matrix,
which has a simple eigenvalue $q^2$, and eigenvalue 0 of multiplicity $q^2 -1$. 
 Thus, $A_D$ has 
eigenvalues $\lambda_0=0$ and $\lambda_1= q$ or $-q$. To argue that we should choose the $+$ sign note that $\text{Tr}A_D=q.$
\end{IEEEproof}

\vspace*{.05in}We now construct a $(1,1)$-recoverable system $X$ over $Q=[q]$ for $q\in\N$. 
According to Lemma \ref{lem:db_d_recov}, $\db([t])$ presents such a system, and if $q=t^2$ for some $t\in\N$, 
then by Lemma \ref{lem:eig_sys_db_graph} the capacity $\ccap(X)=\log_q t=\frac{1}{2}$.
Now suppose that $q$ is not a whole square. As Corollary \ref{cor:sub_db_recov} suggests, we can obtain a 
$(1,1)$-recoverable system by considering a subgraph of a de Bruijn graph. Let $t=\lceil\sqrt{q}\rceil$, $q=t^2-r$, and consider the 
graph $\db([t])$ with the adjacency matrix $A_D$. Taking a subgraph of $\db([t])$ amounts to deleting 
rows and columns from $A_D$. 
The obtained matrix, which we call the \textbf{truncated matrix} $A_{Q}$, is the adjacency matrix of a de Bruijn subgraph that presents a $(1,1)$-recoverable system. This sequence of operations, illustrated in the examples below, will be justified formally in the remainder of this section.

Before proceeding we make one remark. A simple way to construct the truncated matrix, which we largely use, is to delete the last $r=t^2-q$ columns and rows in $A_D$. In what follows we restrict ourselves to alphabets $Q$ of size $q=t^2-r$, where $t=\lceil\sqrt{q}\rceil$ and $r\leq t$. 
Deleting the last $r$ columns from $A_D$ can be interpreted as deleting the outgoing edges of the vertices that correspond to these columns. 
If $r<t$, then upon deleting the last $r$ columns, no all-zeros rows arise. 
At the same time, if $t=r$, deleting the last $r=t$ columns results in $t$ all-zeros rows.
These rows correspond to vertices with no outgoing edges, there is a complication that they are not the last $t$ rows in the matrix. 
While such rows do not contribute to the capacity of the system, we cannot erase them since their incoming edges will point nowhere. 
Thus when $t=r$ we need a different method to delete rows and columns: 
instead of deleting the $t$ last rows and columns, we erase the rows and columns in positions $\mathset{it+(t-1), i=0,1,\dots,t-1}$.n

\begin{example}
\label{ex:dug1}
  {\rm  In this example we construct a $(1,1)$-recoverable system over $Q=[8]$. One possible construction, according to Lemma \ref{lem:db_d_recov}, is to generate $\db([2],3)$ and to use it as a $(1,1)$-recoverable system  with capacity $\frac{1}{3}$ (in fact, it is also  $(1,2)$-recoverable). 
	
	Another way to obtain such a system is to use the procedure described above. Let $t=\ceilenv{\sqrt{q}}=3$ and consider the subgraph of the $9\times 9$ matrix of $\db(3)$ obtained after deleting the last column and last row. We have 
	\[A_Q=\begin{bmatrix}
	1&1&1&0&0&0&0&0 \\
	0&0&0&1&1&1&0&0 \\
	0&0&0&0&0&0&1&1 \\
	1&1&1&0&0&0&0&0 \\
	0&0&0&1&1&1&0&0 \\
	0&0&0&0&0&0&1&1 \\
	1&1&1&0&0&0&0&0 \\
	0&0&0&1&1&1&0&0
	\end{bmatrix}.\] 
	Considering the graph presented by $A_Q,$ we obtain a $(1,1)$-recoverable system with capacity 
	$\log_8 (1+\sqrt{3})\approx 0.483 > \frac{1}{3}$. Here $1+\sqrt 3$ is the largest eigenvalue of the matrix
	$A_Q.$}
\end{example}

\begin{example} {\rm
We now construct a $(1,1)$-recoverable system over $Q=[6]$ using the procedure described above. 
Let $t=\ceilenv{\sqrt{q}}=3$ and consider the matrix $\db(3)$ after deleting rows (and columns) $2,5,8$. 
We have 
	\[A_Q=\begin{bmatrix}
	1&1&0&0&0&0 \\
	0&0&1&1&0&0 \\
	1&1&0&0&0&0 \\
	0&0&1&1&0&0 \\
	1&1&0&0&0&0 \\
	0&0&1&1&0&0 
	\end{bmatrix}.\] 
	Considering the graph presented by $A_Q,$ we obtain a $(1,1)$-recoverable system	with capacity 
	$\log_6 (2)\approx 0.387$. Notice that the last $t-1=2$ columns in $A_Q$ contain zeros. This implies that $A_Q$ corresponds to a graph in which
	the last $t-1$ vertices have no incoming edges. Removing them does not affect the capacity but reduces the alphabet size. Indeed, if we remove those vertices we obtain $\db([2])$ which has capacity $\frac{1}{2}$.}
\end{example}

The truncated matrix $A_Q$ obtained from the above procedure has the following property.
\begin{lemma}
	\label{lem:A_output_alg_str_con}
	Let $Q$ be a finite alphabet of size $q=t^2-r$ with $t=\ceilenv{\sqrt{q}}$. If $r< t$ then the matrix $A_Q$ is an adjacency matrix of a strongly connected graph $G=(V,E,L)$.
\end{lemma}

\begin{IEEEproof} We will prove a stronger statement, namely, that the diameter of the graph $G$
with the adjacency matrix $A_Q$ equals two. In other words, there is a path of length at most $2$ between any two vertices.

Let $q=t^2-r$ where $t=\ceilenv{\sqrt{q}}$ and $r< t$. According to Lemma \ref{lem:shape_db_adj_mat}, every row in the adjacency matrix of $\db(t)$ contains $t$ consecutive ones. The truncated matrix $A_Q$ of order $q$ obtained after the deletion of at most $r$ rows, contains two types of rows, rows with $t$ consecutive ones, and rows with $t-r$ consecutive ones.

If a row contains $t$ consecutive ones, they appear in positions $tj+[t]=\{tj, tj+1,\dots,tj+(t-1)\}$ for some $j<t-1$. 
More accurately, row $j$ contains $t$ consecutive ones in positions $t(j\bmod t)+[t]$ if $j\neq t-1 \bmod t$. 
If $j\bmod t=t-1$ then row $j$ contains $t-r$ consecutive ones in positions $t(t-1)+[t-r]$. 
This implies that for every $j\in [q]$ and every set of $t$ consecutive rows in $A_Q$, there is exactly one row from that set that has $1$ in position $j$. This means that any set of vertices that corresponds to $t$
consecutive rows in the matrix is a dominating set in $G$ (i.e., it connects to all the other vertices in $G$).

A vertex $v$ that corresponds to a row with $t$ consecutive ones is connected to a set of vertices that correspond
to $t$ consecutive rows. Therefore, there is a directed path of length at most 2 from $v$ to any other
vertex in $G$.  
At the same time, every row that contains $t-r$ ones in the last positions, corresponds to a vertex $v$ that
is connected to the last $t-r$ vertices (rows). These $t-r$ vertices form a dominating set in $G$, giving rise to
a path of length 2 from $v$ to any other vertex in $G$.
\end{IEEEproof}

Lemma \ref{lem:A_output_alg_str_con} implies the following statement.
\begin{proposition}
	\label{cor:mat_A_rank_2} Let $Q$ be an alphabet of size $q$ and let $q=t^2-r,$ where $t=\ceilenv{\sqrt{q}}$. Let $A_Q$ be the truncated matrix. 
	For $r< t$, $A_Q$ has at most two non-zero eigenvalues (without accounting for multiplicity).
\end{proposition}

\begin{IEEEproof}
Assume that $r<t$, and let us show that in this case $\rank((A_Q)^2)=2$. In this case the truncation procedure
described above does not give rise to any all-zero rows. The following statements are checked by inspection.
The matrix $A_Q$ contains two types of rows, those with $t$ consecutive ones (located in positions $tj+[t]$ for $j<t-1$) and those with $t-r$ consecutive ones (located in the last $t-r$ positions). 
The row $j$ contains $t-r$ consecutive ones if $j\bmod t=t-1$ and $t$ consecutive ones otherwise. 
Because of this, there are also two different types of columns. A column of the first type contains 
$t$ ones of which every two are separated by $t-1$ zeros. A column of the second type contains
$t-1$ ones of which every two are separated by $t-1$ zeros. 
In columns of the first type there is a single one 
that appears in the last $t-r$ positions, and in columns of the second type the last $t-r$ positions are all 
zeros. The columns of the second type are located in the last $r(t-1)$ positions. The arrangement of the 
rows and the columns is shown in Figure \ref{fig:mat}.
\begin{figure}[t]
\centering
\includegraphics[width=\linewidth]{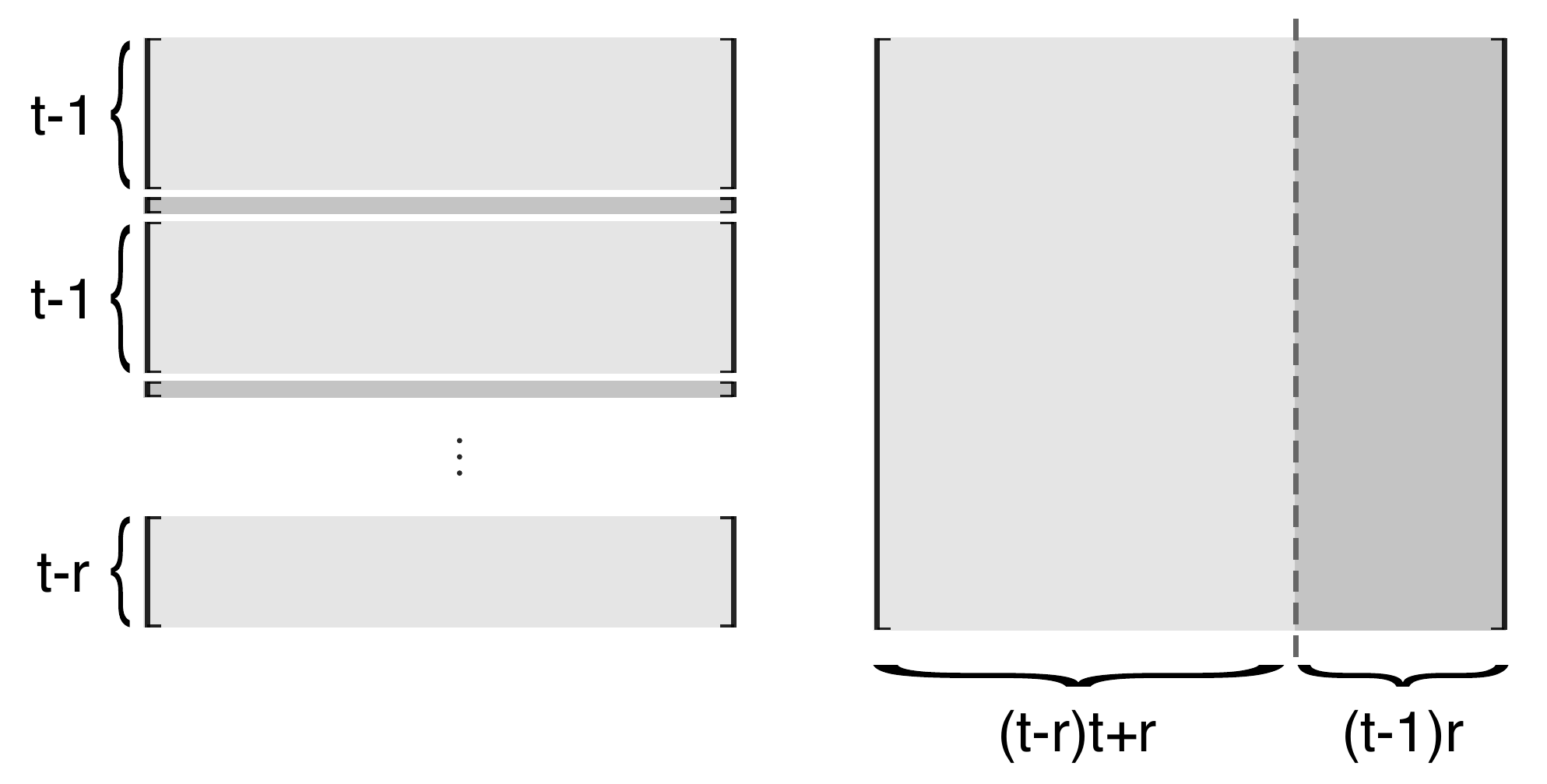}
\caption{The structure of the matrix $A_Q$. The left figure details the arrangement of the rows: each block of size $t-1$ (light gray) contains rows of the first type, and is followed by a single row of the second type (dark gray). Overall there are $t-1$ rows of the second type. The last block is formed of $t-r$ rows of the first type. The right figure describes the structure of the columns. There are $(t-r)t+r$ columns of the first type (light gray) followed by $(t-1)r$ columns of the second type (dark gray).}
\label{fig:mat}
\end{figure}

The structure of the matrix implies that the product of any row with a first-type column is $1$, the product of a first-type row with a second-type column 
is also $1$, and the product of a second-type row with a second-type column is $0$. 
Moreover, the matrix $A_Q^2$ contains rows of only two different kinds, those of all ones and those
with $(t-r)t$ ones followed by $r(t-1)$ zeros. 
Hence $\rank(A_Q^2)=2$ and thus $A_Q^2$ has at most $2$ non-zero eigenvalues, which may or may not be different.
Note that the rank of $A_Q$ can be greater than two; however, an eigenvector of $A_Q$ is also an eigenvector of $A_Q^2$, therefore, the matrix $A_Q$ has at most two eigenvectors with nonzero eigenvalues 
(an eigenvalue $\lambda\ne 0$ of $A_Q$ gives rise to an eigenvalue $\lambda^2$ of $A_Q^2$). This proves the claim for $r<t.$

\end{IEEEproof}

Next we deal with the case $r=t$.
\begin{proposition}
\label{prop:r=t}
    Let $Q$ be a finite alphabet of size $q=t^2-t$ with $t=\ceilenv{\sqrt{q}}$. 
    The Perron eigenvalue of the truncated matrix $A_Q$ is $\lambda=t-1$.
\end{proposition}
\begin{IEEEproof}
Recall that when $r=t$, the matrix $A_Q$ is obtained from $\db([t])$ by deleting rows (and columns) $\mathset{it+(t-1) ~:~ i\in [t]}$. 
The structure of $\db([t])$ suggests the following deletion procedure. Every row of $\db([t])$ contains exactly $t$ ones, and we begin by eliminating all the rows in which the $t$ ones appear at the end, i.e., in positions $(t-1)t+[t]$. There are exactly $t$ such rows with numbers in the set 
$\mathset{it+(t-1) ~:~ i\in [t]}$. Next, delete every column that intersects the last one in some row.
Exactly $t$ columns are thus eliminated. 
Every row of the truncated matrix $A_Q$ contains exactly $t-1$ ones, and so the all-ones vector 
is a positive eigenvector with eigenvalue $t-1$. For a nonnegative matrix, a positive eigenvector corresponds to the largest eigenvalue, and this finishes the proof.
\end{IEEEproof}
\begin{remark} 
  The construction of $A_Q$ in the case $r=t$ generates a $(1,1)$-recoverable system with effective alphabet which is strictly smaller than $q=t^2-t,$ namely it uses only $(t-1)^2$ letters out
the $q$ available letters in $Q$.
\end{remark}

We can now state and prove a lower bound on the maximum capacity of $(1,1)$-recoverable systems. 
\begin{theorem}
	\label{th:exp_cap_alg}
	Let $Q$ be a finite alphabet of size $q=t^2-r$ with $t=\ceilenv{\sqrt{q}}$ and $r\leq t$. Then 
	\begin{align}
	\label{eq:11}
	C_q(1,1)\geq 
	\log_q \parenv{\frac{1}{2}\parenv{t-1+\sqrt{(t-1)^2+4(t-r)}}}. 
	\end{align}
\end{theorem}

\begin{IEEEproof}
	We prove the theorem by explicitly calculating the capacity of the system $X$ presented by the graph $G=(V,E,L)$ with adjacency matrix $A_Q$. 
	First observe that according to Corollary \ref{cor:sub_db_recov}, the system $X$ is indeed a $(1,1)$-recoverable system.  
	
	For $r=t,$ Proposition \ref{prop:r=t} implies that $\lambda=t-1$ and $\ccap(X)=\log_q (t-1)$. 
	For $r<t$, Lemma \ref{lem:A_output_alg_str_con} states that $A_Q$ is strongly connected, which implies that $\ccap(X)=\log_q \lambda,$ where $\lambda$ is the maximal eigenvalue of $A_Q$. Define a vector $v\in\R^q$ as follows.
	For $i\in [q]$ and some real number $\xi$, to be chosen later, let
	\begin{align*}
	v_i&=\begin{cases}
	\xi & \text{ if } i\bmod t =t-1\\
	1 & \text{ otherwise}
	\end{cases}.
	\end{align*}
	Recall that $A_Q$ has two types of rows and refer to Figure \ref{fig:mat} for an illustration. 
We obtain
	\begin{align*}
	(A_Q v)_i&=\begin{cases}
	t-r& \text{ if } i\bmod t =t-1\\
	t-1+\xi & \text{ otherwise}
	\end{cases}.
	\end{align*}
Let us adjust $\xi\in \R$ so that $v$ be an eigenvector of $A_Q$. 
	If $\lambda$ is an eigenvalue, then $A_Q v=\lambda v$ implies the following two relations 
	\begin{align*}
	t-r&=\lambda \xi \\
	t-1+\xi&=\lambda,
	\end{align*}
which are satisfied by
	\begin{align*}
	\xi_{1,2}&=\frac{-(t-1)\pm \sqrt{(t-1)^2+4(t-r)}}{2}, \\
	\lambda_{1,2}&=\frac{(t-1)\pm \sqrt{(t-1)^2+4(t-r)}}{2}.
	\end{align*}
	From Proposition \ref{cor:mat_A_rank_2}, $\lambda_{1,2}$ are the only non-zero eigenvalues. 
Hence, $\lambda_1$ is the maximum eigenvalue which finishes the proof.
\end{IEEEproof}

Even though the de Bruijn graph construction yields $(1,1)$-recoverable systems for all values of $q$,
we were able to compute the resulting capacity only for $q$ that satisfy $t^2-t\le q\le t^2$
for all $t\ge 2.$ 
A quick calculation from \eqref{eq:11} yields that
   $$
C_q(1,1)\ge \log_q(t-1)\ge \frac 12-\frac1{(t-1)\ln q}
   $$
(we used a standard inequality $\ln(1+x)\ge x/(1+x)$ valid for $x>-1$).
Together with the recursion in Sec.~\ref{sec:rec} this yields lower bounds on $C_q(1,1)$ 
for all values of $q\ge 4$. At the same time, the best bound is obtained if the recursion starts 
from the largest whole square smaller than $q$. 

\begin{figure}[th]
\centering
\includegraphics[width=\linewidth]{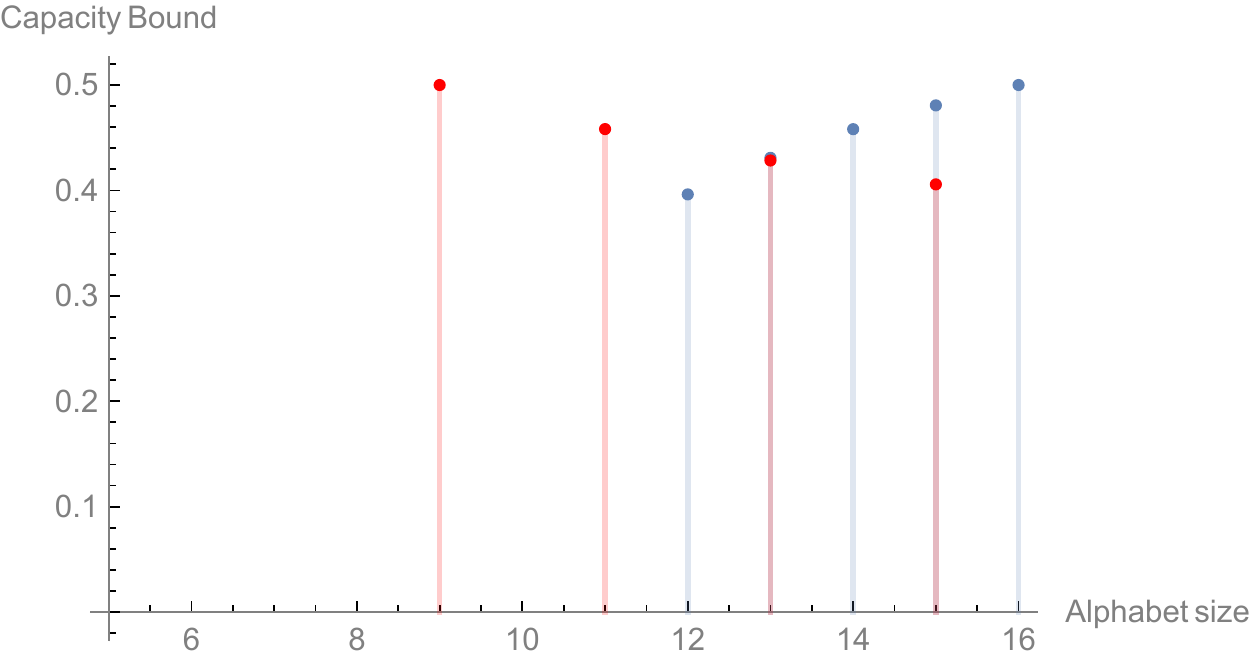}
\caption{The bounds obtained in Sec.~\ref{sec:constructions} for small values of $q$. The bound of 
Theorem \ref{th:exp_cap_alg} is plotted for $q=12,\dots,16$ and the bound of Lemma \ref{lem:lower_bound_m} is plotted for $q=9,11,13,15$. 
To compute the bound of Lemma \ref{lem:lower_bound_m} we started with $q=9$ (with capacity $1/2$) and used it to find the capacity bounds for $q=11,13,15$. }
\label{fig:mat2}
\end{figure}

While the full comparison of the two bounds obtained in this section is cumbersome, in Fig.~\ref{fig:mat2} we show some results for
small values of $q$.

\section{$\epsilon$-recoverable systems}\label{sec:relax}
A natural relaxation of the recoverability requirement is to allow a small fraction of failures in the recovery
of the window of size $k$ in the sequence. In the first part of the section we formalize this concept using
a measure-theoretic definition of recoverable systems.

\subsection{Definition and general properties}
Consider the alphabet $ Q$ together with the discrete $\sigma$-algebra and let $ Q^{\Z}$ be the space of bi-infinite sequences endowed with the product Borel $\sigma$-algebra (under the product topology, $Q^{\Z}$ is compact and metrizable, 
\cite[p.47]{KatokHasselblat1995}.). 
For a sequence $x\in Q^{\Z}$ and a set $F\subseteq \Z$ we denote by $x_{F}$ the projection of $x$ on the 
coordinates in $F$, $x_F\in Q^F$. 
For a word $w\in Q^F$, denote by $\set w_F$ the cylinder set 
\[\set w_F:=\mathset{x\in Q^{\Z} ~:~ x_F=w}.\]
In particular, $\set x_F$ denotes the cylinder set $\set{x_F}_F$ of all sequences in $Q^\Z$ that coincide with $x$ on $F$. For $F=[n]$ we will write $\set w$ instead of $\set w_{[n]}.$ The collection of cylinder sets forms a base of the product topology on $Q^{\Z}$.

For a set $X\subseteq  Q^{\Z}$, we denote by $\cM(X)$ the set of all probability measures on $X.$
We will use the following subsets of $\cM(Q^\Z)$:
\begin{itemize}
\item[$(i)$] $\cM_s(Q^\Z)$ denotes the set of all shift invariant measures;
\item[$(ii)$]  $\cM_r^{(k,l)}(Q^\Z)$, all shift invariant $(k,l)$-recoverable measures (Def.~\ref{def:RS2});
\item[$(iii)$] $\cM_r^{(\epsilon,k,l)}(Q^\Z)$, all $(\epsilon,k,l)$-recoverable measures (Def.~\ref{def:ekl}).
\end{itemize}

For a finite set $F\subset \Z$, we denote by $\mu_F$ a measure on $Q^F$ defined as the $F$-marginal of $\mu$. If  $F=\mathset{i}$ is a singleton, we will write $\mu_i$ instead of $\mu_{\{i\}}$. 

The measure-theoretic counterpart of capacity is called entropy and is defined next. Let $\mu\in\cM_s( Q^{\Z})$ and let $F\subseteq \Z$. Recall that the \textbf{Shannon entropy} 
\[H_{\mu}(F)=-\int \log_{ q} \mu(\set x_F) d\mu(x).\] 
As above, we will write $H_{\mu}(i)$ instead of $H_{\mu}(\mathset{i})$.
The \textbf{entropy} of the measure is defined as 
    \begin{equation}\label{eq:h}
    h(\mu):= \lim_{n\to\infty} \frac{1}{n} H_{\mu}([n])
    \end{equation}
where the existence of the limit is a classic result in ergodic theory, e.g., \cite[Thm.~4.22]{Walters1982}. 

Let $\cA$ be a $\mu$-measurable $\sigma$-algebra on $Q^\Z$. For a set $F\subseteq \Z$ we let 
    \begin{equation}\label{eq:cH}
    H_{\mu}(F | \cA)=-\int \log_{ q} \mu(\set x_F | \cA) d\mu(x)
    \end{equation}
and define the conditional entropy with respect to $\cA$ as 
\[h(\mu | \cA):= \lim_{n\to\infty} \frac{1}{n} H_{\mu}([n] | \cA).\] 

For a set of coordinates $F\subseteq \Z$, we denote by $\sigma(F)$ the $\sigma$-algebra generated by the 
coordinates in $F$ (by all the cylinder sets of the form $\set{w}_{F}$ for some $w$). 

\begin{definition}\label{def:RS2}
	Let $\mu\in\cM_s( Q^{\Z})$ be a shift invariant measure. For $k,l\in\N$, we say that $\mu$ is a $(k,l)$-recoverable measure if 
	\[
	H_{\mu}([k] | \sigma(N_l([k])))=0,
	\]
where $N_l([k])$ is the $l$-neighborhood of the set $[k]$.
	\end{definition}
For a measure $\mu\in\cM_r^{(k,l)}$, shift invariance implies that $H_{\mu}(j+[k] \mid \sigma(j+N_l([k])))=0$
 for every 
$j\in\Z$. Since $X_{[k]}$ can take only a finite number of values (at most $ q^k$), the recoverability property is equivalent to the property that $X_{[k]}$ is measurable with respect to $\sigma(N_l([k]))$.

Let $X_{\cF}\subseteq Q^{\Z}$ be a constrained system. The capacity of $X_\cF$ relates to the notion of entropy via the \textbf{variational principle} \cite[Thm.~8.6]{Walters1982}, \cite[p.181]{KatokHasselblat1995} which states that 
\[\ccap(X_\cF)=\sup_{\mu\in \cM_s(X_\cF)} h(\mu).\] 

To connect the definition of a recoverable system given earlier in the introduction with Definition \ref{def:RS2}, we recall the measure-theoretic approach to the capacity of constrained systems \cite[Sec.3.2.3]{MarRotSie98}.

\begin{construction}[Markov measure]\label{const:markov}
{Let $X_{\cF}$ be a constrained system presented by a graph $G=(V,E,L)$ and let $A_G$ be its adjacency matrix.
Suppose that $G$ is strongly connected. The Perron-Frobenius theorem implies that the left and right
eigenvectors with the largest eigenvalue of $A_G$ are positive (have strictly positive coordinates). Denote
these vectors by $x$ and $y$ and suppose w.l.o.g. that $x^Ty=1$. 
 Define a matrix $P$ by setting 
    $$
    P_{u,v}=\frac{(A_G)_{(u,v)} y_v}{\lambda y_u}, \quad u,v\in V.
    $$
Define a Markov measure $\mu$ on $X_{\cF}$ using $P$ as the transition matrix and the vector $p:=(x_vy_v, v\in V)$
as the starting distribution.  
It suffices to define $\mu$ on the cylinder sets $\set w$. For every path $\gamma=(e_0,\dots,e_{n-1})$ in $G$ with 
$w=L(\gamma)$ simply put
	\begin{align}\label{eq:ma1}
	 \mu(\set w)=p_{v_0} P_{v_0,v_1}\cdots P_{v_{n-1},v_n}
	 \end{align}
where $e_i=(v_i,v_{i+1})$. This completes the construction. }
\end{construction}

The next theorem derives from results on the eigenvalues of the matrix $A_G$, proved in \cite[Thm.~3.9]{MarRotSie98} for irreducible aperiodic matrices and in \cite{LinMar21}, Exercise 4.5.14 without the aperiodicity assumption. The maximality of $h(\mu)$ is also
discussed in \cite[p.~443]{LinMar21}.
\begin{theorem}\label{thm:mme_obtained}
The measure $\mu$ is supported on the set $X_{\cF}$ and is the unique measure for which 
$h(\mu)=\ccap(X_{\cF})=\log \lambda.$ 
\end{theorem}

This theorem 
 together with the variational principle implies that for every $k,l\in\N$, there exists a shift invariant $(k,l)$-recoverable measure $\mu\in\cM_r^{(k,l)}( Q^{\Z})$ with maximum entropy, 
   $$
h_q^*(k,l):=\max_{\mu\in\cM_r^{(k,l)}( Q^{\Z})} h(\mu).
  $$
In addition, any $(k,l)$-recoverable measure $\mu$ that maximizes the entropy is Markov with memory $2l+k-1$. 
Therefore, we will think of a system $X$ as a discrete-time stochastic process 
$X=(\dots, X_{-1},X_0,X_1,\dots)$. 
With this in mind, for a finite set $F=\mathset{n_0,\dots, n_k}\subseteq\Z$, we have
\begin{align}
\label{eq:RV}
\Pr(X_{n_0}=x_{n_0},\dots,X_{n_k}=x_{n_k})
&= \mu\parenv{\set {x_{n_0}\dots x_{n_k}}_F}\\ \nonumber
&= \mu_F\parenv{x_{n_0}\dots x_{n_k}},
\end{align}
and the Shannon entropy takes the form 
   $$
   H_{\mu}(X_{n_0},\dots,X_{n_k})=-\int \log_{q} \mu_F(x) \ d\mu_F(x).
   $$
For a set $G=\mathset{m_0,\dots,m_l}\subseteq\Z$ we denote by 
$H_{\mu}(X_{n_0},\dots,X_{n_k} ~|~ X_{m_0},\dots,X_{m_l})$ the conditional Shannon entropy with respect to the $\sigma$-algebra generated by the pre-image of the random variables $X_{m_0},\dots, X_{m_l}$. 

Let us define the $\epsilon$-recoverability property.
\begin{definition}\label{def:ekl}
	Let $\mu\in\cM_s( Q^{\Z})$ and let $\epsilon>0$. We say that $\mu$ is an \textbf{$(\epsilon,k,l)$-recoverable measure} if for every $\alpha,\beta\in Q^l$, 
       \begin{equation}\label{eq:ekl}H_{\mu}(X_{l+[k]} ~|~ X_{[l]}=\alpha, X_{l+k+[l]}=\beta)\leq \epsilon.
\end{equation}
We say that $\mu$ is $\epsilon$-recoverable if there exist some numbers $k,l$ such that $\mu$ is $(\epsilon,k,l)$-recoverable. 
\end{definition}

The problem that we consider in this section is to characterize $(\epsilon,k,l)$-recoverable measures with maximum entropy. At this moment 
however, it is not clear that this problem is well posed. Indeed, the support of an $\epsilon$-recoverable measure is not necessarily a 
constrained system since such a measure in  principle can assign a positive probability to any word (cylinder set) $w\in Q^\ast.$
At the same time, a constrained system is defined by a set of forbidden patterns $\cF\subseteq  Q^*$ which therefore should have
probability zero. For this reason, it is not immediately clear if there exists an $(\epsilon,k,l)$-recoverable measure $\mu$ that maximizes 
the entropy $h(\mu)$ in the class of such measures. 
In the following we show that such a measure indeed exists and describe some of its properties.

\begin{proposition}
The set $\cM_r^{(\epsilon,k,l)}$ contains a measure whose entropy attains the 
supremum $\sup_{\mu\in\cM_r^{(\epsilon,k,l)}(Q^\Z)}h(\mu)$.
\end{proposition}

\begin{IEEEproof}We observe the following.
\begin{enumerate}
    \item \label{en:1} For every finite set $F\subset \Z$, the coordinate restriction map $\pi_F$ given by $\mu\mapsto \mu_F$ is continuous. 
    Indeed, for every $w\in Q^F$ we have 
    \begin{align*}
    \abs{\pi_F(\mu_1)(w)-\pi_F(\mu_2)(w)} 
    =\abs{\mu_1(\set{w})-\mu_2(\set{w})}.
    \end{align*}
    Thus, $\abs{\pi_F(\mu_1)(w)-\pi_F(\mu_2)(w)}\leq \|\mu_1-\mu_2\|_{TV}$ for every cylinder set, which implies that 
    $\|\pi_F(\mu_1)-\pi_F(\mu_2)\|\leq \|\mu_1-\mu_2\|_{TV}$.

\item \label{en:2} The Shannon entropy $H_q(\mu_F)$ is a weak$^*$ continuous function of the argument (discrete distribution) $\mu_F$. 
\item \label{en:3} The conditioning operation $\mu(\cdot)\mapsto \mu(\cdot \mid A),$ where $A$ is an event, is weak$^*$ continuous as a mapping of measures. Below we denote this operation by $\mathsf{Con}_A$.
\end{enumerate}
From \ref{en:2}), for every $\alpha,\beta\in Q^l$, the inverse image of the closed set $[0,\epsilon]$ under the Shannon entropy gives a closed set of (conditional) measures on $k$-tuples, i.e., the set 
\[ M_{\alpha,\beta}:=H_{X_{l+[k]} \mid X_{[l]}=\alpha, X_{l+k+[l]}=\beta}^{-1}\parenv{[0,\epsilon]}\subseteq \cM(Q^{k})\]
is closed. 
Let $A$ be the event 
\[\mathset{x\in Q^{[2l+k]} ~:~ x_{[l]}=\alpha, x_{l+k+[l]}=\beta}.\]
From \ref{en:3}) we obtain that for every $\alpha,\beta\in Q^l,$ the set $\mathsf{Con}_{A}^{-1}(M_{\alpha,\beta})\subseteq \cM(Q^{2l+k})$ is also closed. This implies that the finite intersection 
     \[
     M:= \bigcap_{\alpha,\beta\in Q^l}\mathsf{Con}_{A}^{-1}(M_{\alpha,\beta})\subseteq \cM(Q^{2l+k})
     \] 
is also closed.
Upon setting $F=[2l+k]$, \ref{en:1}) implies that 
$\pi_F^{-1}(M)\subseteq \cM(Q^{\Z})$ 
is a closed set. 
Since $\cM_s(Q^{\Z})$ is closed and compact, taking the intersection $\cM_s(Q^{\Z})\cap \pi_F^{-1}(M)$ we obtain a closed (and hence compact) set. This is exactly the set of all $(\epsilon,k,l)$-recoverable measures. The shift operator $T$ is continuous, which implies that the function $\mu\to h(\mu)$ is affine \cite[Thm.~8.1]{Walters1982}. Finally, an affine function on a compact set attains its supremum.\end{IEEEproof}

We conclude that the problem stated above after Definition \ref{def:ekl} is well posed. Therefore, we introduce
the following notation: let
   $$
   h^*_q(\epsilon,k,l):=\max_{\mu\in\cM_r^{(\epsilon,k,l)}(Q^\Z)}h(\mu).
   $$
The next lemma argues that this maximum is attainable by a Markov measure, which also
implies that the entropy $h^*_q(\epsilon,k,l)$ is finite \cite[Thm. 4.27]{Walters1982}.

\begin{lemma}
	\label{lem:mme_markov}
	Let $\epsilon>0$ and let $\mu$ be an $(\epsilon,k,l)$-recoverable measure with $h(\mu)=h^*_q(\epsilon,k,l)$. 
	Then there exists a Markov measure $\nu$ with memory $2l+k-1$ such that $h(\nu)=h^*_q(\epsilon,k,l)$.
\end{lemma}

\begin{IEEEproof}
We start by explaining the idea of the proof. To shorten the notation, we limit ourselves to the case 
$k=l=1$, but the general proof follows the same arguments. The proof follows the known construction of measure extensions \cite{Sch1985,Gold2017}. The key point is that the 
$(\epsilon,k,l)$-recoverable property is completely determined by the marginal distribution $\mu_{[2l+k]}$. 
We will construct a memory-$2$ Markov measure $\nu$ with the same distribution over $ Q^3$ as $\mu$, sometimes termed
the 2nd order Markov approximation of $\mu.$ 
Since $\nu_{[3]}=\mu_{[3]},$ we can claim that $\nu$ is an $(\epsilon,1,1)$-recoverable measure, and show that
$\nu$ maximizes the entropy.

	For every $m\in\N,$ define a measure $\nu_m$ over $ Q^m$ as follows. For $1\leq m\leq 3$ and for every $w\in Q^m$ define $\nu_m(\set w)=\mu(\set w)$. For $m=4$ and every $w=(w_0,w_1,w_2, w_3)\in Q^m$ define \[\nu_m(\set w)=\mu(\set{w_0w_1w_2})\mu(\set{w_1w_2w_3} | \set{w_1w_2})\]
	and for $w=(w_0,\dots, w_{m-1})\in Q^m$, $m>4$ define $\nu_m$ recursively in a similar fashion:
	\begin{multline}\label{eq:mar_ex}
	    \nu_m(\set w)=\nu_{m-1}(\set{w_{[m-1]}}) \\ 
	    \times\mu(\set{w_{m-3}w_{m-2}w_{m-1}}|\set{w_{m-3}w_{m-2}}).
	\end{multline} 
	Since $\nu_m$ is defined for every $m\in\N$, by the Kolmogorov extension theorem there exists a (unique) measure $\nu$ over $ Q^{\Z}$ with $\nu_m$ as its marginals. Specifically, $\nu_{[3]}=\mu_{[3]},$ which implies that $\nu$ is $(\epsilon,1,1)$-recoverable. Since $\mu$ is shift invariant, so is $\nu,$ and by construction $\nu$ is Markov with memory $2$. To finish the proof, we show that $\nu$ 
maximizes the entropy over all shift invariant measures that coincide with $\mu$ on $ Q^3$. 
	First notice that for $m\leq 3$ we have $H_{\mu}([m])=H_{\nu}([m])$, where $H_{\mu}([m])=H_\mu(\set{x_{[m]}})$, for instance, is the Shannon entropy of the cylinder set. For $m>3$ we have 
	\begin{align*}
	H_{\nu}([m])&=H_{\nu_m}([m]) \\
	&=H_{\nu_{m-1}}([m-1])+H_{\mu}([3] \mid [2]) \\ 
	&=H_{\nu_{m-1}}([m-1])+H_{\mu}([3])-H_{\mu}([2])
	\end{align*}
where the first equality follows from \eqref{eq:mar_ex} and shift invariance of $\mu$, and the second 
uses the chain rule for entropies $H_{\mu}([3])=H_{\mu}([2])+H_{\mu}([3] ~|~ [2])$. Here the notation $H_{\mu}([3] \mid [2])$ means $H_{\mu}([3] \mid \sigma([2])),$ i.e., conditioning on the $\sigma$-algebra generated by the coordinates $0,1$ in the sequence $w$ (see \eqref{eq:cH}).
	We obtain 
	\begin{align}
	\label{eq:nu_ent}
	H_{\nu}([m])-H_{\nu_{m-1}}([m-1])=H_{\mu}([3])-H_{\mu}([2]).
	\end{align}
Now	consider $H_{\mu}([m])$ for some $m\in\N$. Using again the chain rule, we obtain 
	$$
	H_{\mu}([m])=H_{\mu}([m-1])+H_{\mu}( \{m-1\} | [m-1])
	$$
and since $H_{\mu}( \{m-1\} | [m-1])\le H_{\mu}( \{m-1\} | \mathset{m-3,m-2})$ 
(conditioning on a sub-$\sigma$-algebra reduces entropy), we have 
	\begin{align*}
H_{\mu}([m])\leq
H_{\mu}([m-1])+H_{\mu}( \{m-1\} | \mathset{m-3,m-2}).
   \end{align*}
	Since $\mu$ is shift invariant, we obtain 
	\begin{align}
	\label{eq:ent1}
	H_{\mu}([m])\leq H_{\mu}([m-1])+H_{\mu}( \{2\}| [2]).
	\end{align}
We also have 
	\[H_{\mu}([3])= H_{\mu}([2])+H_{\mu}(\{2\}| [2]).\] 
	Subtracting this equality from \eqref{eq:ent1}, we obtain
	\begin{align*}
	H_{\mu}([m])- H_{\mu}([m-1]) \leq H_{\mu}([3]) -H_{\mu}([2]).
	\end{align*}
Taken together with \eqref{eq:nu_ent}, this inequality implies that $\nu$ maximizes the entropy over all shift invariant measures whose length-$3$ marginals agree with $\mu$. 
\end{IEEEproof}

\vspace*{.1in}
Next we show that $h^*_q(\epsilon,k,l)$ is obtained when the relaxation is exploited to its full extent. This fact is formally stated in the next theorem, whose proof draws 
in part on the ideas of \cite{BurSte1994}.
We need to introduce additional elements of notation. Let $\Z^{-}$ denote the set of all negative integers and let $\cC$ ($\cC^-$) denote the $\sigma$-algebra generated by $\Z$ ($\Z^-$), i.e., by cylinders $\set w_F$ with $F\subseteq \Z$ or $F\subseteq \Z^-$, respectively.
Since we consider only shift invariant measures, we have $h(\mu)=H_{\mu}(0 ~|~ \cC^{-})$ \cite[Eq.~(15.19)]{Geo1988}. Using shift invariance and the chain rule, we obtain that for every $n\in\N$, 
\begin{align}
\label{eq:ent_as_mar}
h(\mu)=\frac{1}{n} H_{\mu}([n] ~|~ \cC^-).
\end{align}
We remark that while this general result applies to all shift invariant measures, we use it for Markov measures. We could limit the conditioning in \eqref{eq:ent_as_mar} to $2l+k-1$ 
coordinates in the past, but the arguments below do not depend on this.
\begin{theorem}
	\label{th:final1}
	Fix $\epsilon>0$ and let $\mu$ be an $(\epsilon,k,l)$-recoverable measure with $h(\mu)=h^*_q(\epsilon,k,l)$. Then there exist $\alpha,\beta\in Q^l$, such that 
	$$
	H_{\mu}(X_{l+[k]}~|~ X_{[l]}=\alpha, X_{l+k+[l]}=\beta)=\epsilon.
	$$
\end{theorem}
\begin{IEEEproof} 
Assume toward a contradiction that for every $\alpha,\beta\in Q^l$, $H_{\mu}(X_{l+[k]}~|~ X_{[l]}=\alpha, X_{l+k+[l]}=\beta)<\epsilon$. 
Let $\alpha,\beta$ be words such that $\sum_{w\in Q^k} \mu_{[2l+k]}(\alpha w \beta)>0$ and that  
$H_{\mu}(X_{l+[k]}~|~ X_{[l]}=\alpha, X_{l+k+[l]}=\beta)$ is maximal. 
We will show that 
$H_{\mu}(X_{l+[k]}~|~ X_{[l]}=\alpha, X_{l+k+[l]}=\beta)=\epsilon$.

Let $\mathbi{p}=(p_0,\dots,p_{ q^k-1})$ be a probability distribution on $Q^k$ with 
$H_q(\mathbi{p}):=-\sum_i p_i\log_q p_i=\epsilon$. Using $\mu$ and $\mathbi{p},$ we will construct a measure $\nu$ that has higher 
entropy than $\mu.$ Define $\nu$ independently on segments of length $m$ in $\Z,$ where the value of $m$ will be specified later. 
To sample a sequence $x\in Q^\Z$ from $\nu$,
choose $x\in  Q^{\Z}$ according to $\mu$ and then, if $x_{[l]}=\alpha, x_{l+k+[l]}=\beta$ then 
replace $x_{l+[k]}$ with a word chosen according to $\mathbi{p}$. Repeat this procedure independently for every 
$j\in\Z$: if $x_{jm+[l]}=\alpha, x_{jm+l+k+[l]}=\beta,$ then replace $x_{jm+l+[k]}$ with a word chosen according to $\mathbi{p}$. 
The resulting measure $\nu$ is invariant under $T^m$ but not necessarily shift invariant. 
To make it shift invariant, consider
$\nu_s=\frac{1}{m}\sum_{i\in [m]}T^i(\nu)$ where $T(\nu)$ is the push-forward of $\nu$, defined by 
$T(\nu)(\cdot)=\nu\parenv{T^{-1}(\cdot)}$.

Let us show that $\nu_s$ is an $(\epsilon,k,l)$-recoverable measure. It suffices to show that for every $w_1,w_2\in Q^l$, $H_{\nu_s}(X_{l+[k]}~|~ X_{[l]}X_{l+k+[l]}=w_1 w_2)\leq\epsilon$. Note that for $l+k\leq i\leq m-2l-k$ we have 
\begin{multline*}
\nu(X_{i+l+[k]}\mid X_{i+[l]}X_{i+l+k+[l]}=w_1 w_2)
\\ =\mu(X_{i+l+[k]}\mid X_{i+[l]}X_{i+l+k+[l]}=w_1 w_2).
\end{multline*}
By definition, we have 
\begin{multline*}
T^i (\nu)(X_{i+l+[k]}\mid X_{i+[l]}X_{i+l+k+[l]}=w_1 w_2) \\
=\nu \parenv{X_{l+[k]}\mid X_{[l]}X_{l+k+[l]}=w_1 w_2}.
\end{multline*}
Since the conditional entropy is continuous and since for every $w_1,w_2\in Q^l$, 
$H_{\mu}(X_{l+[k]}~|~ X_{[l]}X_{l+k+[l]}=w_1 w_2)<\epsilon$, 
we can choose $m$ large enough such that
$H_{\nu_s}(X_{l+[k]}~|~ X_{[l]}X_{l+k+[l]}=w_1 w_2)\leq\epsilon$. Generally, the value of $m$ depends on $w_1,w_2,$ but there are finitely many possibilities to choose them, so
it is possible to choose a large enough, but finite, $m$ such that
$\nu_s$ is an $(\epsilon,k,l)$-recoverable measure.

Since $\mu$ maximizes the entropy, we have that 
	\begin{align}
	\label{eq:fi1}
	h(\nu_s)\leq h(\mu).
	\end{align}
	
Next we define a partition $\cA^m=\cA^m_1\cup\cA^m_2$ 
of $Q^\Z$ according to the values of $x$ in the coordinates in $[m].$ Let $r=m-2l-k$ and note
that $[m]=\cup_{i=1}^4 J_i,$ where $J_1= [l], J_2=\{l+[k]\},J_3=\{l+k+[l]\},J_4=\{2l+k+[r]\}.$
The class $\cA^m_1$ is formed of cylinder sets $\set{x_{[m]}}$ such that $x_{J_1}\ne \alpha$ or 
$x_{J_3}\ne\beta.$ The class $\cA^m_2$ is formed of cylinder sets $\set{x_{[m]}}$
such that $x_{J_1}= \alpha, x_{J_3}=\beta$ and $x_{J_4}=w,$ where $w\in Q^r.$ 
Formally,
  \begin{gather*}
  \cA_1^m=\bigcup_{u\in Q^{[m]},\; u_{J_1}u_{J_3}\neq \alpha\beta}\set u\\
   \cA_2^m= \bigcup_{w\in Q^r} A_2^m(w), \quad A_2^m(w)=\bigcup_{v\in Q^k}\set{\alpha v\beta w}.
   \end{gather*}
Thinking of the system as a sequence of random variables \eqref{eq:RV}, let us rewrite
\eqref{eq:ent_as_mar} as
	\begin{equation}\label{eq:mu1}
	h(\mu)= \frac{1}{m}H_{\mu}(X_{[m]} ~|~ \cC^-).
	\end{equation}
Now consider a function $f$ that takes $x_{[m]}$ to its corresponding part in $\cA^m$. 
Namely, if $x_{J_1}\neq \alpha$ or $x_{J_3}\neq \beta,$ then $f$ takes $x_{[m]}$ to itself. Otherwise, $f$ takes $x_{[m]}$ to the part defined by $x_{J_1},x_{J_3},x_{J_4}$, i.e., $f$ does not distinguish between $m$-words for which $x_{J_1}=\alpha, x_{J_3}=\beta ,x_{J_4}=w$. 

Clearly, we have 
	\[H_{\mu}(f(X_{[m]}) ~|~ \cC^-,X_{[m]})=0,\]
	which together with the chain rule implies that 
	\begin{align*}
	H_{\mu}(X_{[m]},f(X_{[m]}) ~|~ \cC^-)= H_{\mu}(X_{[m]} ~|~ \cC^-).
	\end{align*}
	Therefore, 
\begin{align}
		&H_{\mu}(X_{[m]} ~|~ \cC^-) =  H_{\mu}(X_{[m]},f(X_{[m]}) ~|~ \cC^-) \nonumber \\ 
	&= H_{\mu}(f(X_{[m]}) ~|~ \cC^-) + H_{\mu}(X_{[m]} ~|~ \cC^-, f(X_{[m]})). \label{eq:fi2}
 \end{align}
For two partitions $\cG_1,\cG_2$ of a set, their refinement is
 defined as $\cG_1\vee \cG_2=\{ g_1\cap g_2: g_1\in\cG_1,g_2\in\cG_2\}.$ 
 Consider the partition $\vee_{j\in\Z} T^{jm}\parenv{\cA^m}$ and let $\cA$ be the $\sigma$-algebra generated by it. 
 Note that $\cA$ is a sub-$\sigma$-algebra of $\cC$ under which $X_{[m]}$ and $f(X_{[m]})$ cannot be distinguished. Further, let $\cA^-$ be the sub-$\sigma$-algebra of $\cA$ obtained by the restriction of $\cA$ to $\Z^-.$ 
 Noticing that $\cC^-$ is a refinement of $\cA^-$, we obtain
	\begin{align}
	\nonumber
	H_{\mu}(f(X_{[m]}) ~|~ \cC^-) &{\leq} H_{\mu}(f(X_{[m]}) ~|~ \cA^-) \\ \nonumber
	&{=}H_{\nu}(f(X_{[m]}) ~|~ \cA^-) \\ \label{eq:fi3}
	&{=} H_{\nu}(f(X_{[m]}) ~|~ \cC^-),
	\end{align}
where on the second line we use the fact that the distribution of $f(X_{[m]})$ is the same under $\mu$ and $\nu$. To justify \eqref{eq:fi3}, 
we notice that $\cA$ is the Borel $\sigma$-algebra of the system after applying $f$ and it is generated by $\cA^m$. 
Therefore, according to the Kolmogorov-Sinai theorem \cite{Sinai1959}, \cite[Thm. 4.17]{Walters1982}, the entropy of $f(X_{[m]})$ does not 
change if instead of $\cA^-$ we condition on $\cC^-.$ 
Intuitively, the last equality follows since $\cA$ contains all the information needed to describe $f(X_{[m]})$ explicitly.

Substituting \eqref{eq:fi3} into \eqref{eq:fi2}, we obtain 
	\begin{align*}
	H_{\mu}(&X_{[m]} ~|~ \cC^-) \nonumber \\
	&\leq H_{\nu}(f(X_{[m]}) ~|~ \cC^-) +
	H_{\mu}(X_{[m]} ~|~ \cC^-, f(X_{[m]}))\nonumber\\
    &\le H_{\nu}(f(X_{[m]}) ~|~ \cC^-) + 	H_{\mu}(X_{[m]} ~|~ f(X_{[m]})). 
	\end{align*}
By the construction of $\nu$ and the assumption about $p$ 
we have 
    \begin{align}
H_{\mu}(X_{[m]} \mid  f(X_{[m]}))&\le H_{\nu}(X_{[m]} \mid f(X_{[m]})) \nonumber \\  
&=H_{\nu}(X_{[m]} \mid \cC^-, f(X_{[m]})) \label{eq:fi5}
      \end{align}
(the past is independent of $X_{l+[k]}$ conditional on $f(X_{[m]})$),
      which implies 
	\begin{multline*} 
	H_{\mu}(X_{[m]} ~|~ \cC^-)\leq 
	H_{\nu}(f(X_{[m]}) ~|~ \cC^-) \\+
	H_{\nu}(X_{[m]} ~|~ \cC^-, f(X_{[m]})).
	\end{multline*}
Using 	\eqref{eq:mu1}, we obtain that
  \begin{multline}\label{eq:mu}
  h(\mu)\le 
  \frac{1}{m}(H_{\nu}(f(X_{[m]}) ~|~ \cC^-) \\+
	H_{\nu}(X_{[m]} ~|~ \cC^-, f(X_{[m]}))).
	\end{multline}
	
Since $\nu$ is invariant with respect to $T^{m}$ (i.e., invariant under a subgroup of $\Z$), monotonicity of the entropy implies that the limit in the definition of $h(\nu)$ exists (see \eqref{eq:h}). Since $\nu_s$ is shift invariant with respect to $\Z$, we conclude that $h(\nu_s)=h(\nu).$ Using the chain rule in \eqref{eq:ent_as_mar}, we obtain that
	\begin{align*}
	h(\nu_s)= 
	\frac{1}{m}H_{\nu}(f(X_{[m]}) | \cC^-)  
	+\frac{1}{m}H_{\nu}(X_{[m]} | \cC^-, f(X_{[m]})).
	\end{align*}
Therefore, on account of \eqref{eq:fi1}, all the inequalities in 
\eqref{eq:fi3}-\eqref{eq:mu} are equalities. 
Noticing that if $X_{[l]}X_{l+k+[l]}\ne \alpha\beta,$ then $H_{\mu}(X_{[m]} \mid f(X_{[m]}))=0,$ we conclude from \eqref{eq:fi5} that
    \begin{align*}
  H_{\mu}(&X_{l+[k]} \mid X_{[l]}=\alpha,X_{l+k+[l]}=\beta) \\
    &=H_{\nu}(X_{l+[k]} \mid X_{[l]}=\alpha,X_{l+k+[l]}=\beta) \\
    &=\epsilon.
    \end{align*}
This contradicts the assumption on $\mu$.
\end{IEEEproof}

\begin{remark}{\rm
An informal justification of the claim of Theorem \ref{th:final1} is as follows. Let $\mu$ be a measure on $Q^{\Z}$ such that for every $\alpha,\beta\in Q^l$, the conditional entropy $H_{\mu}(X_{l+[k]} | X_{[l]}=\alpha, X_{l+k+[l]}=\beta)<\epsilon$. Now consider a measure $\nu$ that is a mixture of $\mu$ with an i.i.d. uniform distribution on $Q^{\Z}$, viz., $\nu= (1-r)\mu + r\eta$ where $r\in [0,1]$ and $\eta$ is a measure on $Q^{\Z}$ such that the entries are distributed 
i.i.d uniform. The value of $r$ can be chosen such that $\nu_{[2l+k]}$ is arbitrarily close to $\mu_{[2l+k]}$ in the total variation distance. By continuity of entropy we can choose $r>0$ such that 
   $$
   \max_{\alpha,\beta\in Q^l}{H_{\nu}(X_{l+[k]} | X_{[l]}=\alpha, X_{l+k+[l]}=\beta)}<\epsilon.
   $$
  Since the entropy is strictly concave, $h(\nu)>h(\mu),$ contradicting the maximality of $\mu$. 
A similar approach is used below to generate $(\epsilon,1,1)$-recoverable systems by increasing the entropy of $(1,1)$-recoverable systems. }
\end{remark}

The following proposition implies that measures with a more relaxed recovery requirement have higher
metric entropy.
\begin{proposition} Let $k,l\ge 1$ and $\epsilon_1>\epsilon_2>0$, then
    $$
    h^\ast_q(\epsilon_1,k,l)>h^\ast_q(\epsilon_2,k,l).
    $$
\end{proposition}
\begin{IEEEproof} Let $\mu_i,i=1,2$ be such that $h(\mu_i)=h^\ast_q(\epsilon_i,k,l).$
Clearly, $h(\mu_1)\ge h(\mu_2)$ because $\mu_2$ is $(\epsilon_1,k,l)$-recoverable. If this inequality holds with equality, then by Theorem~\ref{th:final1} there are $\alpha,\beta\in Q^l$ such that
  $$
  H_{\mu_2}(X_{l+[k]}~|~ X_{[l]}=\alpha, X_{l+k+[l]}=\beta)=\epsilon_1>\epsilon_2,
  $$
a contradiction.
\end{IEEEproof}

The actual recovery procedure of a $k$-tuple from its $l$-neighborhood is rather 
straightforward and reduces to taking the $k$-word that has the largest conditional 
probability with respect to the measure $\mu.$ This statement will be made more formal once we prove Lemma \ref{lem:uniq_sym_high_prob} in the next section.

\subsection{Constructing $(\epsilon,k,l)$-recoverable measures}
To construct recoverable measures we will rely on $(k,l)$-recoverable systems.
It is intuitive that for small $\epsilon,$ the capacity $h^\ast(\epsilon,k,l)$ should approach the capacity of recoverable systems $C_q(k,l).$
In this section we formalize this intuition, transforming $(k,l)$-recoverable systems into $\epsilon$-recoverable measures. This approach has its limitations in the sense that
it works only for small values of $\epsilon.$ 

Below we use the entropy function $H_q(x):=-x\log_q x-(1-x)\log_q (1-x),$ where $q\ge 2$ is an integer. We start
with a well-known property of the binary entropy.
\begin{lemma}
	\label{lem:low_bound_ent} For all $x\in[0,1]$
$$
H_2(x)\ge 4x(1-x).
$$
\end{lemma}
\begin{IEEEproof} Consider the function $f(x)=\frac{H_2(x)}{x(1-x)}.$ It is immediate that $f'(x)\le 0$
for $x\in[0,1/2],$ and so $f(x)\ge f(1/2)=4,$ or $H_2(x)\ge 4x(1-x).$ By symmetry around 1/2 we can claim
this inequality also for $1/2<x\le 1.$
\end{IEEEproof}

In the next lemma we state a simple property of recoverable measures, namely that small conditional entropy implies that the symbols can be recovered with high probability. In the statement, $\set{\alpha w\beta}$ 
refers to a cylinder set with
$\alpha,\beta\in Q^l$ and $w\in Q^k.$ 

\begin{lemma}
	\label{lem:uniq_sym_high_prob}
Let $X$ be distributed according to a measure $\mu\in\cM_s( Q^{\Z})$ that is $(\epsilon,k,l)$-recoverable, and suppose that $0<\epsilon< \frac{2}{q^k}$. Then for every $\alpha,\beta\in Q^l$ with $\sum_{w'\in Q^k}\mu(\set{\alpha w'\beta})>0$, there exists a unique $w\in Q^k$ such that $\Pr(X_{l+[k]}=w ~|~ X_{[l]}=\alpha, X_{l+k+[l]}=\beta)\geq 1-\frac{\epsilon}{2}$.
\end{lemma}

\begin{IEEEproof}
We begin with the case $k=l=1$ and $ q=2$. Since $\mu$ is $(\epsilon,1,1)$-recoverable, we have that $X_1$ 
conditioned on $X_{0}=\alpha, X_2=\beta$ has a Bernoulli$(p)$ distribution for some $p\in[0,1],$ such
that $H_{\mu}(X_1 ~|~ X_{0}=\alpha, X_2=\beta)=H_2(p)\leq \epsilon$. By Lemma \ref{lem:low_bound_ent} we have 
	\[\epsilon\geq H_2(p)\geq 4p(1-p).\]
which implies that
	\[p\geq \frac{1+\sqrt{1-\epsilon}}{2}\qquad \text{ or } \qquad p\leq \frac{1-\sqrt{1-\epsilon}}{2}.\]
	Since $\sqrt{1-\epsilon}\geq 1-\epsilon$ we have 
	\[p\geq 1-\frac{\epsilon}{2} \qquad \text{ or } \qquad p\leq \frac{\epsilon}{2}.\] 
This implies that there is a symbol among $\{0,1\}$ with probability $\ge 1-\epsilon/2.$ 

	Now let $ q>2$ and $k,l\in\N$ and let us reduce this general case to the binary case. Fix $\alpha,\beta\in  Q^l$ and define the mapping $f: Q^k\to \mathset{0,1}$ as follows. Let $w\in Q^k$ be such that 
	\begin{align}
	\Pr(X_{l+[k]}&=w | X_{[l]}=\alpha, X_{l+k+[l]}=\beta) \nonumber\\ 
	&\geq \Pr(X_{l+[k]}=w' | X_{[l]}=\alpha, X_{l+k+[l]}=\beta) \nonumber\\ 
	&>0 	\label{eq:prob1}
	\end{align}
	for every $w'\neq w$. Notice that such a word $w$ exists since $\sum_{w'\in Q^k}\mu(\set{\alpha w'\beta})>0$. Let $w'\in Q^k$ and define 
	$$
	f_w(w')={\mathbbm{1}}(w'\ne w),
	$$
so $f_w(X_{l+[k]})$ can be regarded as a binary RV. Thus, conditional on the event that 
$X_{[l]}=\alpha,X_{l+k+[l]}=\beta,$ the RV $f_w(X_{l+[k]})$ is distributed according to
Bernoulli$(p)$ for some $p\in [0,1]$. For definiteness, put $p:=\Pr(X_{l+[k]}=w | X_{[l]}=\alpha, X_{l+k+[l]}=\beta)$, where $w$ is chosen according to \eqref{eq:prob1}.

We have
	\begin{align*}
	H_{\mu}(f_w(X_{l+[k]}) &~|~ X_{[l]}=\alpha,X_{l+k+[l]}=\beta) \\
	&\leq H_{\mu}(X_{l+[k]} ~|~ X_{[l]}=\alpha,X_{l+k+[l]}=\beta)
	\end{align*}
	which implies
	$$
	H_{\mu}(f_w(X_{l+[k]}) ~|~ X_{[l]}=\alpha,X_{l+k+[l]}=\beta)\leq \epsilon.
	$$
	
	From the binary case analysis we obtain that
	\[p\geq 1-\frac{\epsilon}{2} \qquad \text{ or } \qquad p\leq \frac{\epsilon}{2}.\] 
	Recall that $w$ is chosen according to \eqref{eq:prob1}. Since 
	\begin{align*}
	1=\sum_{w'\in Q^k} \Pr(X_{l+[k]}=w'| X_{[l]} 
	=\alpha,X_{l+k+[l]}=\beta)\le q^kp 
	\end{align*}
This relation together with the assumption that $\epsilon< \frac{2}{ q^k}$ implies that $p\geq 1-({\epsilon}/{2})$.
\end{IEEEproof}

\begin{remark}\label{rem:410}
  {\rm   1. Lemma \ref{lem:uniq_sym_high_prob} implies that for every $\alpha,\beta\in Q^l$, there is a unique $w_{\alpha,\beta}\in Q^k$ such that $\mu(\set{\alpha w_{\alpha,\beta} \beta})$ is close to $1$. 
Observe that the small entropy constraint translates into a bound on the error probability of recovery. Define
  $$
   p_e(X):= 
   \max_{\alpha,\beta\in Q^l} P(X_{l+[k]}=w_{\alpha,\beta}|X_{[l]}=\alpha,X_{l+k+[l]}=\beta)
   $$
to be the maximum error probability, then the condition of $(\epsilon,k,l)$-recoverability clearly implies that $p_e\le \epsilon/2.$

2. Define the set 
    $$
\cF=\bigcup_{\alpha,\beta\in Q^l}\mathset{\alpha w \beta ~:~ w\neq w_{\alpha,\beta}}.
    $$
    The constrained system $X_{\cF}$ with the set of forbidden words $\cF$ is 
$(k,l)$-recoverable, and moreover, if $\mu$ is a measure for which $h(\mu)=h^*_q(\epsilon,k,l)$ then $\ccap(X_\cF)=C_q(k,l).$ Indeed, if not, then there is a 
constrained system $Y$ with $\ccap(Y)>\ccap(X_\cF).$ For a sufficiently small $\epsilon$
the construction of $\cF$ and continuity of entropy imply that $h(\mu)$ is arbitrarily close to $\ccap(X_\cF).$ Since $\mu$ is such that $h(\mu)$ attains the value $h_q^\ast(\epsilon,k,l),$ and since $Y$ gives rise to a measure $\nu$ with $h(\nu)=\ccap(Y),$ we obtain that $h(\mu)\ge \ccap(Y).$ This yields a contradiction, proving our claim.
    
    }
\end{remark}

Remark \ref{rem:410}.2 raises the question whether this relation also applies in reverse direction, i.e., whether it is possible to obtain an $(\epsilon,k,l)$-recoverable measure with maximum entropy from a $(k,l)$-recoverable system with maximum capacity. 
In the remainder of this section we establish a partial result toward the resolution of this question. Namely, given a $(k,l)$-recoverable system and $\epsilon>0$, we construct an $(\epsilon,k,l)$-recoverable measure and calculate its entropy, obtaining a lower bound on $h_q^\ast(\epsilon,k,l)$. 

\subsubsection{Informal description of the construction.} The $(\epsilon,k,l)$-recoverable measure will 
be constructed by perturbing a Markov measure associated with a maximum-capacity 
deterministic $(k,l)$-recoverable system. 
We start with a graph that presents a $(k,l)$-recoverable system $X_\cF$ with maximum capacity $C_q(k,l).$ 
Since the recoverability property is defined by words of length $2l+k$, we use a presentation $G=(V,E,L)$ of $X_\cF$ such that vertices in $V$ correspond to words of length $2l+k$ over $Q$. 
This system is equivalently described by a Markov measure defined in \eqref{eq:ma1}. 
 Denoting it by $\rho$, let $P^{\rho}$ and $p^{\rho}$ be the matrix of transition probabilities and the stationary distribution of $\rho$, respectively.

To describe the perturbation, it will be convenient
to switch to a different Markov measure, which we proceed to define.
Let $Y$ be the system obtained from $X_{\cF}$ by considering non-overlapping subwords of length $2l+k$ (also called the $(2l+k)$th higher power of $X_{\cF}$ \cite{LinMar21}). We can view $Y$ as a system over the alphabet $Q^{2l+k}$.
Define a Markov measure $\mu$ on $Y\subseteq (Q^{2l+k})^{\Z}$ obtained from $\rho$ as follows. 
The state set of $\mu$ is the same as the states of $\rho$. The matrix of transition probabilities of $\rho$ is obtained
as $P^{\mu}=(P^{\rho})^{2l+k}$, and the stationary distribution is unchanged, i.e., $p^{\mu}=p^{\rho}$. 
In other words, $\mu$ is obtained from the $(2l+k)$ power of 
the graph $G$ (the graph $G^{2l+k}$ on $V$ whose edges correspond to paths 
of length $2l+k$ in $G$). 
Consider a system $Z\subseteq (Q^{2l+k})^{\Z}$ obtained by passing $Y$ through a conditional distribution, given by
a stochastic matrix  of order $2l+k$ with entries  
   \begin{multline*}
    W(w_0aw_2|w_0w_1w_2)= 
    \frac\delta{q^k-1}\mathbbm{1}(a\in Q^k\backslash\{w_1\})\\+\bar\delta\mathbbm{1}(a=w_1),
    \end{multline*}
where $\delta>0$ and $\bar\delta:=1-\delta.$
This matrix can be also viewed as a memoryless communication channel on $Q^{2l+k}.$
As a result of this operation, we obtain a new system, which we denote by $Z$. This system is presented by a graph $D$ whose
set of vertices is $V$ together with the new vertices of the form $(w_0 a w_2), a\in Q^k\backslash\{w_1\}.$ 
The distribution on the sequences of the system $Z$ is obtained as $\nu(\set w)=\mathbb{E}_\mu(W(\set x|\cdot)),$ where $w\in Q^{2l+k},$
which can be extended to the cylinder sets of $(Q^{2l+k})^\Z$ by independence. The construction of $\nu$ is explained in detail below.

We finally use $\nu$ to construct an $(\epsilon,k,l)$-recoverable measure $\eta$ over $Q^{\Z}$ similarly to the procedure that appears in the proof of Lemma \ref{lem:mme_markov}. 

\begin{figure}[h]
\centering 
\includegraphics[width=3in]{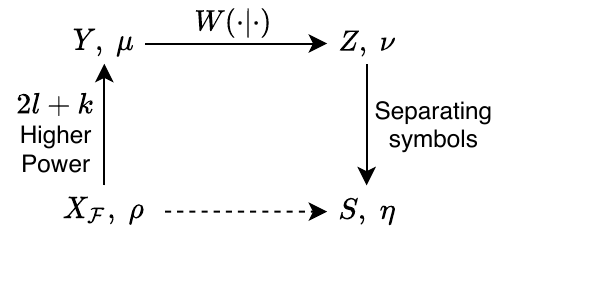}
\vspace*{-.3in}\caption{The procedure of constructing $\eta\in\cM_s(Q^{\Z})$. From $X_{\cF}$ (with the Markov measure $\rho$) we construct $Y$ with a measure $\mu\in\cM_s((Q^{2l+k})^{\Z})$ using the $(2l+k)$th higher power of $X_{\cF}$. We then obtain $Z$, distributed according to $\nu$, by passing the states 
of $Y$ through a memoryless channel. In the last step we obtain the system $S$ over the alphabet $Q$, together with the measure $\eta\in\cM_s(Q^{\Z})$. }\label{fig:proc}
\end{figure}

\subsubsection{Formal description} 
\begin{figure}[ht]
\centering \includegraphics[width=3in]{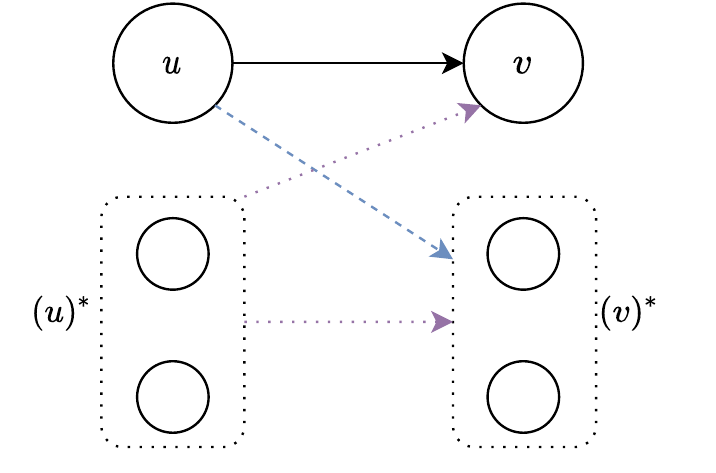}
\vspace*{-.1in}\caption{Construction of the graph $D$}\label{fig:Puv}
\end{figure}

We proceed to implement this plan. The following sequence of steps transforms a graph that represents a $(k,l)$-recoverable system to a graph that supports $\epsilon$-recoverability.
\begin{construction}\label{const:D}
\begin{enumerate}
\item\label{step1} Let $X_{\cF}$ be a $(k,l)$-recoverable system with $\ccap(X_{\cF})=C_q(k,l)$, presented by a graph $G'$.
The vertices of $G'$ correspond to $(2l+k-1)$-tuples of symbols in $Q$. 

\item\label{step2} Consider the set 
$$
\cF':=\mathset{w\in Q^{2l+k+1} : \exists u\in\cF \text{ s.t. } u\prec w}
$$
and take $G=(V,E,L)$ to be the graph that presents $X_{\cF'},$ so that its vertices correspond
to words of length $2l+k$ over $Q$.
Note that $X_{\cF'}=X_{\cF}$, so the graph $G$ gives another presentation of the system $X_\cF.$

\item\label{step3} Consider the graph $$G^{2l+k}=(V(G^{2l+k}),E(G^{2l+k}),L(G^{2l+k}))$$ with the same set of vertices as $G$. Two vertices in $G^{2l+k}$ are connected by an edge if and only if there is a path of length $2l+k$ connecting them in $G$ (including self-loops). The label of the edge $(v_1,v_2)\in E(G^{2l+k})$ is given by the label of the vertex $v_2.$ 

\item\label{step4} Construct a graph $D=(V(D),E(D),L(D)).$ Given a vertex $u=(u_0 u_1 u_2),\; u_0,u_2\in Q^l,\; u_1\in Q^k$ we write $u_a:=(u_0 a u_2),$ where $a\in Q^k$ is some $k$-tuple.  The set of vertices
$V(D)$ is formed of $V(G^{2l+k})$ and all the vertices of the form $u_a=(u_0 a u_2), a\in Q^k\backslash\{u_1\}$ where $u=(u_0 u_1 u_2)\in V(G^{2l+k}).$ The set $E(D)$ contains all the edges of the form $(u_a,v_b),$ where  $(u,v)\in E(G^{2l+k}).$ 
\end{enumerate}
\IEEEQED
\end{construction}
The details of the construction are illustrated in Example~\ref{ex:cont} in the Appendix, and
the sequence of steps of the construction is shown schematically in Figure \ref{fig:proc}.
To add details to Step~\ref{step4}, note that if a vertex $u\in V(G^{2l+k})$ corresponds to a triple
$u=(u_0 u_1 u_2),\; u_0,u_2\in Q^l,\; u_1\in Q^k,$ then no other vertex in $V(G^{2l+k})$ is of the form $(u_0 a u_2),$
where $a\ne u_1.$ In constructing $V(D)$ we add all the vertices of this form to $V(G^{2l+k})$,
denoting the set of vertices that arise from $u\in V(G^{2l+k})$ by $(u)^\ast$; see Fig.~\ref{fig:Puv}.

The graph $G^{2l+k}$ presents a constrained system over $Q^{2l+k}$ which we denote by $Y$. 
It is evident that $Y$ can be obtained from $X_{\cF}$ by reading non-overlapping subwords 
of length $2l+k$ in the words $x\in X_{\cF}$, 
and $X_{\cF}$ can be obtained from $Y$ by inverting this operation. 
It is a known fact that $\ccap(Y)=\ccap(X_{\cF})$ (see \cite[Prop. 3.13]{MarRotSie98} or 
\cite[Exercise 4.1.5]{LinMar21},
where this equality uses a different logarithm base and has a slightly different form). Accordingly, the system $Z,$ presented by the graph $D$, is also over the alphabet $Q^{2l+k},$ and it will be used to construct an $(\epsilon,k,l)$-recoverable system $S$ over $Q.$

\vspace*{.05in}
Let us define transition probabilities for the graph $D$. 
Theorem \ref{thm:mme_obtained} implies that there exists a shift invariant Markov
measure $\mu$ supported on $Y$ and such that $h(\mu)=\ccap(Y).$ Denote by $P^\mu$ the 
matrix of transition probabilities on $V(G^{2l+k})$ derived from $\mu.$ 
By shift invariance, the marginal distribution of $\mu$ on any one 
coordinate equals the stationary distribution of $P^\mu.$ 
Denote this stationary distribution by $p^{\mu}$.

Let $\epsilon>0$ and let $\delta \in [0,(q-1)/q]$ be such that 
\begin{align}
\label{eq:cent}
H_q(\delta)+\delta \log_q 
(q^k-1)=\epsilon.
\end{align}  

Given a pair of vertices $(u,v)\in V(D)^2,$ let
  \begin{equation}\label{eq:Puv}
  P_{uv}=\begin{cases}
   \bar\delta P_{uv}^\mu &\text{if }(u,v)\in E(G^{2l+k})\\[.05in]
   \frac\delta{q^k-1} P_{u\bar v}^\mu &\text{if }u,\bar v\in V(G^{2l+k}); v\in (\bar v)^\ast\\[.05in]
   \bar\delta P_{\bar uv}^\mu &\text{if } u\in (\bar u)^\ast; \bar u, v\in V(G^{2l+k})\\[.05in]
   \frac\delta{q^k-1} P_{\bar u\bar v}^\mu &\text{if }
   \bar u,\bar v\in V(G^{2l+k}); \\ 
   &\quad u\in (\bar u)^\ast,v\in (\bar v)^\ast
   \end{cases}
\end{equation}
(recall that $\bar\delta=1-\delta$). Referring to the example in the appendix (Fig.~\ref{fig:sub2}), the solid edges correspond to line 1, the dashed edges to line 2, and the dotted
edges to lines 3,4 in \eqref{eq:Puv}, respectively.

By construction, $P$ is a stochastic matrix.
This follows because the system $Z$ presented by the graph $D$ can be viewed as the output of a memoryless channel. Moreover, since $P$ is irreducible, the stationary distribution exists and is unique. Together with the fact that for vertices $u\neq v$ in $V(G^{2l+k})$, the sets $(v)^*\cap(u)^*$ are disjoint, this implies the following result.

\begin{lemma}\label{lemma:sd}
Define a vector $p^{\nu}$ of dimension $q|V(G^{2l+k})|$ by
     \begin{equation}\label{eq:pv}
         p^{\nu}_u=\begin{cases} \bar\delta p^{\mu}_u&\text{if }u\in V(G^{2l+k})\\
         \frac{\delta}{q^k-1}p^{\mu}_{\bar v} &\text{otherwise,}
         \end{cases}
     \end{equation}
where $\bar v\in V(G^{2l+k})$ is the unique vertex such that $u\in (\bar v)^\ast$. 
Then $p^{\nu}$ is a probability vector and $p^{\nu}=p^{\nu} P.$
\end{lemma}
 
The stationary distribution $p^{\nu}$ together with $P$ gives rise to a shift invariant
measure $\nu$ on a set $Z\subseteq (Q^{2l+k})^\Z$ of bi-infinite sequences over $Q^{2l+k}.$ 
For symbols $w_i\in Q^{2l+k},\; i\in [n],$ the probability of the cylinder set $\set{w_0 \dots w_{n-1}}$ is given by $p^{\nu}_{w_0} P_{w_0 w_1}\cdots P_{w_{n-2}w_{n-1}}$.

Our next goal is to construct an $(\epsilon,k,l)$-recoverable measure $\eta$ from $\nu$.
Let $Z=(\dots,Z_{-1},Z_0, Z_1,\dots )$ be a system distributed according to $\nu.$  
As the first step, let us transform sequences of $(2l+k)$-tuples $z\in (Q^{2l+k})^{\Z}$ into 
sequences $s\in Q^{\Z}$. Next, we construct a Markov approximation measure $\eta$ to the measure $\nu$ 
using the procedure similar to the one used in the proof of Lemma~\ref{lem:mme_markov},
and such that $h(\eta)\geq h(\nu).$ 
Recall that $\nu$ is a $1$-step Markov measure that corresponds to a system over the alphabet $Q^{2l+k}$. 

\begin{lemma}
    The measure $\eta\in \cM(Q^{\Z})$ obtained from $\nu$ is shift invariant.
\end{lemma}

\begin{IEEEproof}
The fact that $\eta$ is shift invariant follows from the shift invariance of $\rho$, and hence of $\mu$. 
As explained in \eqref{eq:mar_ex}, $\eta$ is obtained from the stationary distribution $\nu$ on $Q^{2l+k}$ by extending $\nu$ to finite sequences. 
Since $\eta$ is a $(2l+k)$ order Markov approximation of $\nu$, we may take the state space of $\eta$ to be $(2l+k)$-tuples. 
Since $\eta$ has the same $(2l+k)$-tuples distribution as $\nu$, both $\eta$ and $\nu$ have the same stationary distribution. Any Markov process with initial stationary distribution is stationary, i.e., shift invariant.
 \end{IEEEproof}

Let us show that  $\eta$ is an $(\epsilon,k,l)$-recoverable measure. 
We will in fact show more, namely that the entropy condition is satisfied with equality.

\begin{proposition}
    Let $\eta$ be a measure on $Q^{\Z}$ obtained from $\nu$ and denote by $S=(\dots,S_{-1},S_0,S_1,\dots)$ the system that is distributed according to $\eta$. For every $\alpha,\beta\in Q^l$, we have 
    $$
    H_{\eta}(S_{l+[k]}\mid S_{[l]}=\alpha, S_{l+k+[l]}=\beta)=\epsilon.
    $$
\end{proposition}

\begin{IEEEproof}
By the definition of $\eta$, for every $s=(s_0 \dots s_{2l+k-1})\in Q^{2l+k}$, we have $\nu(\set s])=\eta(\set s)$. 
Note that since $\nu$ is constructed from the shift invariant measure $\mu$, if $s\in Q^{2l+k}$ with $\eta(\set s)>0$, then $\nu(\set s)>0$. 
Let $\alpha,\beta\in Q^l$ and define a probability vector $\mathbi{p}=(p_a)$ by setting
    $$
    p_a=\frac{\nu(\set{\alpha a \beta})}{\sum_{b\in Q^k} \nu(\set{\alpha b\beta})}, \quad  a\in Q^k.
    $$
Let us show that $H_q(\mathbi{p})=\epsilon.$ Let $c\in Q^k$ be such that $\alpha c\beta\in V(G^{2l+k})$, then $\sum_{b\in Q^k} \nu(\set{\alpha b\beta})=\mu(\set{\alpha c\beta})$ by Lemma~\ref{lemma:sd}. We obtain that $p_c=\bar\delta,$ and for $a\neq c$, $p_a=\frac{\delta}{q^k-1}$. From \eqref{eq:cent} we conclude that 
$H_q(\mathbi{p})=\epsilon$, which was to be shown. 
\end{IEEEproof}

\vspace*{.1in}Our next goal is to calculate the entropy $h(\nu),$ where we recall that
$\nu$ is constructed using a $(k,l)$-recoverable measure $\mu$ over $(Q^{2k+l})^{\Z}$. In the proof we use a standard expression
for the entropy of a Markov chain $Z$ with transition probabilities $P$ and stationary distribution $p^{\nu}$:
    $$
    h(\nu)=-\sum_{u\in V}p^{\nu}_u\sum_{v\in V}P_{uv}\log P_{uv}
    $$
(see \cite[Thm.~4.27]{Walters1982}).

\begin{proposition}\label{th:fin33}
    We have 
    $$
    h(\eta)\geq h(\nu)=h(\mu)+\frac{1}{2l+k}\epsilon.
    $$
\end{proposition}

\begin{IEEEproof}
The inequality follows since $\eta$ is the $(2l+k)$th Markov approximation of $\nu,$ so we only need to compute $h(\nu)$. Let $(Z,\nu)$ be the system obtained by Construction \ref{const:D}. 
We compute logarithms to the base $q^{2l+k}$ and omit the base below in the proof. Using \eqref{eq:pv} and \eqref{eq:Puv}
and recalling that $V(D)\subset V(G^{2l+k})$, we
split the sum on $u,v$ into four parts as follows:
 \begin{align*}
-h(\nu)&= \sum_{u\in V(D)}p^{\nu}_u \sum_{v\in V(D)}P_{uv}\log  P_{uv}\\ 
&= \sum_{u\in V(G^{2l+k})}p^{\nu}_u \sum_{v\in V(G^{2l+k})}P_{uv}\log  P_{uv} \\
&\quad + \sum_{u\in V(G^{2l+k})}p^{\nu}_u \sum_{v\notin V(G^{2l+k})} 
P_{uv}\log  P_{uv} \\ 
&\quad + \sum_{u\notin V(G^{2l+k})}p^{\nu}_u \sum_{v\in V(G^{2l+k})}P_{uv}\log  P_{uv} \\
&\quad + \sum_{u\notin V(G^{2l+k})} 
p^{\nu}_u \sum_{v\notin V(G^{2l+k})} P_{uv}\log  P_{uv}.
\end{align*}
Evaluating the first sum, we obtain 
\begin{align*}
  &\sum_{u\in V(G^{2l+k})}p^{\nu}_u \sum_{v\in V(G^{2l+k})}P_{uv}\log  (P_{uv}) \\ 
  &=\sum_{u\in V(G^{2l+k})}\bar{\delta}p^{\mu}_u \sum_{v\in V(G^{2l+k})}\bar{\delta}P^{\mu}_{uv}\log  (\bar{\delta}P^{\mu}_{uv}) \\
  &= \bar{\delta}^2\sum_{u\in V(G^{2l+k})}p^{\mu}_u \sum_{v\in V(G^{2l+k})}P^{\mu}_{uv}\log  (\bar{\delta}P^{\mu}_{uv}) \\
  &= -\bar{\delta}^2 h(\mu)+\bar{\delta}^2\sum_{u\in V(G^{2l+k})}p^{\mu}_u\sum_{v\in V(G^{2l+k})}P^{\mu}_{uv}\log \bar{\delta}\\
  &= -\bar{\delta}^2 h(\mu)+\bar{\delta}^2\log  \bar{\delta}.  
\end{align*}
Similarly, the second sum evaluates to
\begin{align*}
   & \sum_{u\in V(G^{2l+k})}p^{\nu}_u \sum_{v\notin V(G^{2l+k})} P_{uv}\log  (P_{uv}) \\
   &= \sum_{u\in V(G^{2l+k})}\!\!\!\bar{\delta}p^{\mu}_u \;\Bigg(\sum_{v\in V(G^{2l+k})} \frac{(q^k-1)\delta}{q^k -1} \cdot \\
   & \qquad \qquad \qquad \qquad \times P^{\mu}_{uv}\log  \parenv{\frac{\delta}{q^k-1}P^{\mu}_{uv}}\Bigg) \\ 
    &= \!\!\!\sum_{u\in V(G^{2l+k})}\!\!\!\bar{\delta}p^{\mu}_u \!\!\!\sum_{v\in V(G^{2l+k})} \!\!\!\delta P^{\mu}_{uv}\log  
    \parenv{\frac{\delta}{q^k-1}P^{\mu}_{uv}}\\ 
    &= -\bar{\delta}\delta h(\mu) + \bar{\delta}\delta\log \frac{\delta}{q^k-1},
\end{align*}
where the first equality follows from \eqref{eq:Puv} and the fact that there are $q^k-1$ additional vertices in $V(D)$ for every vertex in $V(G^{2l+k})$.

The third and fourth sum are easily found to be
\begin{align*}
    \sum_{u\notin V(G^{2l+k})}\!\!\!p^{\nu}_u \!\!\!\sum_{v\in V(G^{2l+k})}\!\!\!P_{uv}\log  (P_{uv})
    =-\delta\bar{\delta} h(\mu)+
    \delta\bar{\delta}\log  \bar{\delta},
\end{align*}
and 
\begin{multline*}
    \sum_{u\notin V(G^{2l+k})} p^{\nu}_u \sum_{v\notin V(G^{2l+k})} P_{uv}\log  P_{uv} 
   \\ = -\delta^2 h(\mu)+\delta^2 \log  \frac{\delta}{q^k-1}.
\end{multline*}

Collecting the terms and using $\bar{\delta}=1-\delta$, we obtain 
\begin{align*}
    h(\nu)&=\parenv{\bar{\delta}^2+2\bar{\delta}\delta+\delta^2}h(\mu)- (\bar{\delta}^2 +\delta\bar{\delta})\log  \bar\delta \\ 
    &\quad -(\delta^2+\delta\bar{\delta})\log  \delta +(\delta+\bar{\delta})\delta\log  \parenv{q^k-1} \\
    &= (\delta+\bar{\delta})^2 h(\mu) -\bar{\delta}\log  \bar{\delta} - \delta \log  \delta \\
    &\hspace*{.5in} +\delta\log  \parenv{q^k-1}\\ 
    &= h(\mu)+H_{q^{2l+k}}(\delta)+ \delta\log  \parenv{q^k-1}.
\end{align*}
Using \eqref{eq:cent} we obtain 
   \begin{align*}
 h(\nu)&=h(\mu)+\frac{1}{2l+k} \parenv{H_q(\delta)+\delta\log_{q} \parenv{q^k-1}}\\
 &=h(\mu)+\frac{\epsilon}{2l+k},
   \end{align*}
which finishes the proof.
\end{IEEEproof}

The next theorem forms one of the main results of this section and is an immediate consequence of Theorem \ref{th:fin33}.
\begin{theorem} 
\label{th:mainthm1}
    Let $Q$ be a finite alphabet and let $\epsilon>0$. Then 
    \begin{equation}\label{eq:>}
    h^*_q(\epsilon,k,l)\geq C_q(k,l)+\frac{1}{2l+k}\epsilon.
    \end{equation}
\end{theorem}

At this point one may wonder if the construction presented in this section 
is optimal, i.e., it yields the maximum value of entropy \eqref{eq:>} for $\epsilon\leq 2/q^k$. The answer is negative, as shown in the next example\footnote{In \cite{EliBar2021} we incorrectly claimed that for small $\epsilon$ 
the value in \eqref{eq:>} is the maximum entropy of a $(\epsilon,k,l)$-recoverable system. }.

\begin{example}
{\rm Let $q=2$ (the binary alphabet) and let $k=l=1$.  In this case $2/q^k=1$, so we take some
$\epsilon< 1.$
According to Theorem \ref{th:mainthm1} and Proposition \ref{ex:1}, applying the construction above to the system with maximum capacity yields an 
$(\epsilon,1,1)$-recoverable system with entropy $C_2(1,1)+\epsilon/3\leq 0.42+1/3\leq 0.8$. 

Now consider a system $Y$ formed of all i.i.d. sequences with entries chosen according to 
$\nu \sim \text{Bernoulli}(\frac{\epsilon}{2})$.
Notice that the entropy of the entire system is 
\[h(\nu)=H_2\parenv{\frac{\epsilon}{2}}\]
where $H_2$ as before is the binary entropy function. By continuity of the entropy function, we can choose $\epsilon<1$ such that $h(\nu)\geq 0.9$. 

In order to connect the system $Y$ to the construction presented above, one may think of $Y$ as the system obtained by the construction when applied to 
the zero-entropy system containing the single sequence $(\dots 000\dots)$. The system $Y$ is an $(H\parenv{\frac{\epsilon}{2}},1,1)$-recoverable system with entropy $\geq 0.9$. }
\end{example}

%
%
%

\section{Concluding remarks}\label{sec:conclusion}
\subsection{Relation to storage codes}\label{sec:storage}
In this section we discuss the connection between recoverable systems and storage coding. As noted above, storage codes in graphs were defined in several independent papers \cite{Maz2015,Riis2007,ShanDim2014}. We will use the definition of \cite{Maz2015}, which is phrased in terms close to our work. Let $G=(V,E)$ be a finite graph and assume that $V=\{v_0,v_1,\dots,v_{n-1}\}.$ The neighborhood of a vertex $v\in V$ is an ordered collection of vertices
$N(v)\subseteq V$ such that $(v,u)\in E$ if and only if $u\in N(v).$
A storage code $\cC_G$ with recovery graph $G=(V,E)$ and $V=[n]$ is a subset of $ Q^n$ together with $n$ deterministic recovery functions $f_v: Q^{N(v)}\to  Q$ 
such that $c_v=f_v(c_{u},u\in N(v))$ for every $c=(c_0,\dots,c_{n-1})\in\cC_G$ and every vertex $v\in V$. In words, a
storage code is a set of vectors of length $n$ such that the symbol in location $i$ can be recovered from the symbols in locations specified by the neighbors of $i$ 
in $G$. 

The main problem studied in the literature is the capacity, or largest cardinality of a storage code for a given graph. 
Formally, the {\em capacity} of a storage code is defined as 
$M_{q}(G)=\frac{1}{n} \log_{ q} |\cC_G|$, and the absolute capacity is defined as $M (G)=\sup_{ q\ge 2} M_{ q}(G)$. The definitions in earlier works usually do not include the normalization $1/n,$ which we have added to better relate it to the notation adopted in this paper. The capacity is known for odd cycles $C_n$ (and equals 1/2) \cite{BKL2013,MazMcgVor2019}, and there are multiple bounds in the literature for other types of graphs \cite{BKL2013,Cameron2016,MazMcgVor2019,BargZemor2021}.

Before we describe the relation between storage codes and recoverable systems, let us make the following observation. Assume that $X_{\cF}$ is a $(k,l)$-recoverable system. A sequence $x\in X_{\cF}$ is said to have period $r$ for some $r\in \N$ if $x_i=x_{i+r}$. 
Let $\per_n (X_{\cF}) $ denote the set of words in $X_{\cF}$ with period $n$. 
It is known that the growth rate of the number of periodic sequences approaches the capacity of the system, namely:
    \begin{align*}
        \limsup_{n\to\infty} \frac{1}{n}\log_q | \per_n (X_{\cF}) | 
        =\ccap(X_{\cF}).
    \end{align*}
\cite[Thm. 4.3.6]{LinMar21}. This implies a construction of storage codes for a family of graphs, which we illustrate 
for the cycle $C_n.$ 

Suppose that $X_{\cF}$ is a $(1,1)$-recoverable system, and note that a periodic sequence corresponds to the labels of a cycle in the graph $G$ that presents it. The collection of $n$-words obtained from the cycles forms a storage code for $C_n.$ We state this fact in the next obvious proposition.
\begin{proposition}
	\label{lem:rec_sys_cyc_repa_codes}
Let $X_{\cF}$ be a $(1,1)$-recoverable system. Then $\per_n (X_{\cF})$ is a storage code for $C_n$. 
\end{proposition}

The set $\per_n(X_{\cF})$ is shift invariant, i.e., if a word $w=(w_0,\dots,w_{n-1})\in \per_n(X_{\cF})$ then also the cyclic shift  $(w_1,\dots,w_{n-1},w_0)\in \per_n(X_{\cF})$. 
This means that there is a single recovery function $f$ such that every symbol $w_i$ can be obtained by applying $f$ with the arguments $w_{i-1},w_{i+1}$ (with indices mod $n$). 

For the case $q=2$ it is possible to obtain an explicit formula for the size $|\per_n(X_{\cF})|$. 
\begin{proposition} \label{lem:num_codewords_C_n}
Let $X_{\cF}$ be a binary $(1,1)$-recoverable system with maximum capacity constructed in Proposition \ref{ex:1}. Then 
	$$
	|\per_n(X_{\cF})|=\lambda_1^n+\lambda_2^n+\lambda_3^n
	$$
where $\lambda_1,\lambda_2,\lambda_3$ are the roots of $x^3-x-1$. 
\end{proposition}

\begin{IEEEproof}
Let  $A$ be the adjacency matrix of the graph $G$ given in Figure \ref{fig:G}, i.e.,
	\[A=\begin{bmatrix}
	0 & 1 & 1 \\
	0 & 0 & 1 \\
	1 & 0 & 0
	\end{bmatrix}.\] 
The number of length-$n$ cycles in $G$ equals the trace of $A^n$, $|\per_n(X_{\cF})|=\Tr(A^n)$. For convenience we denote $a_n=(A^n)_{1,1}$, $b_n=(A^n)_{2,2}$, $c_n=(A^n)_{3,3}$, thus $|\per_n(X_{\cF})|=a_n+b_n+c_n$. 
Clearly, we have the following recursion: $a_n=a_{n-2}+a_{n-3}$ with initial values $a_0=1,a_1=0,a_2=1$. Further, 
from the form of the matrix $A$ it is readily seen that $b_n=(A^{n-1})_{2,1}$, and we obtain a recursion $b_n=b_{n-2}+b_{n-3}$ with initial values $b_0=1,b_1=0,b_2=0$. Similarly, $c_n=c_{n-2}+c_{n-3}$ with initial values $c_0=1,c_1=0,c_2=1$. 

Altogether, the sum $a_n+b_n+c_n$ satisfies a recursion  $z_n=z_{n-2}+z_{n-3}$ with initial conditions $z_0=3,z_1=0,z_2=2$. The sequence $(z_n)_n$ is known as the Perrin sequence \cite{AdaSha1982} in which the $n$th number is given by $z_n= \lambda_1^n+\lambda_2^n+\lambda_3^n$ where $\lambda_1,\lambda_2,\lambda_3$ are the roots of $x^3-x-1$ (sequence A001608 in OEIS; see http://oeis.org/A001608) . 
\end{IEEEproof}
Notice that $\ccap(X_{\cF})=\log \lambda_1\approx 0.4057$. Moreover, $\lambda_1>1$, and $|\lambda_2|,|\lambda_3|<1$, hence for large $n$ we obtain $|\per_n(X_{\cF})|\approx 2^{0.4507 n}$. 

Let us stress again the differences between this result and the results in earlier works such as \cite{BKL2013,Cameron2016,MazMcgVor2019}: we analyze the size of the storage code for $C_n$ for a given alphabet $q$,
and we require the same recovery function for each vertex $v$, while the earlier works consider the supremum $\sup_q$ 
and allow different functions $f_v.$

The property of $C_n$ that makes this construction possible is cyclic automorphism of the graph $C_n.$ 
A more general family of graphs with a cyclic automorphism is {\em circulant graphs}, i.e., cycles with chords. 
A {\em circulant graph} is a graph $G=(V,E)$ with $V=(v_0,\dots,v_{n-1})$ such 
that the graph $TG$ obtained by relabeling the vertices $v_i\mapsto v_{i+1}$ (modulo $n$) is isomorphic to $G$.
The bound in Proposition \ref{lem:num_codewords_C_n} extends to this case without difficulty.
Namely, if $X_{\cF}$ is a $(k,l)$-recoverable system, where $l$ is large enough to account for all the neighbors of the
vertex, then $\per_n(X_{\cF})$ provides a storage code for $G$. 

\begin{figure*}[!t]
\hspace*{.4in}\begin{subfigure}{0.45\textwidth}	\centering
    \includegraphics[width=\linewidth]{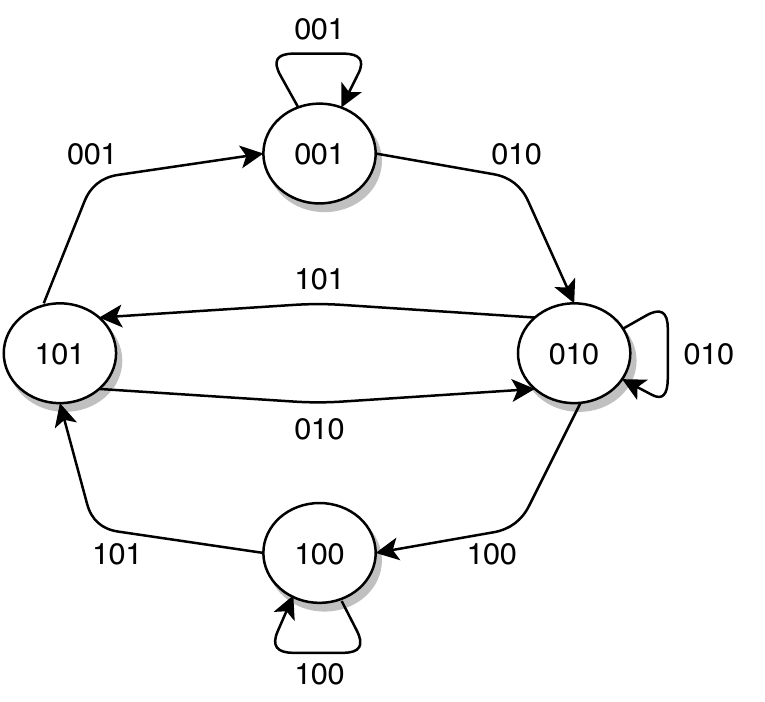} 
\caption{}
\label{fig:sub1}
\end{subfigure}
\hspace{0.3cm}
\begin{subfigure}{0.45\textwidth}
\centering
\includegraphics[width=\linewidth]{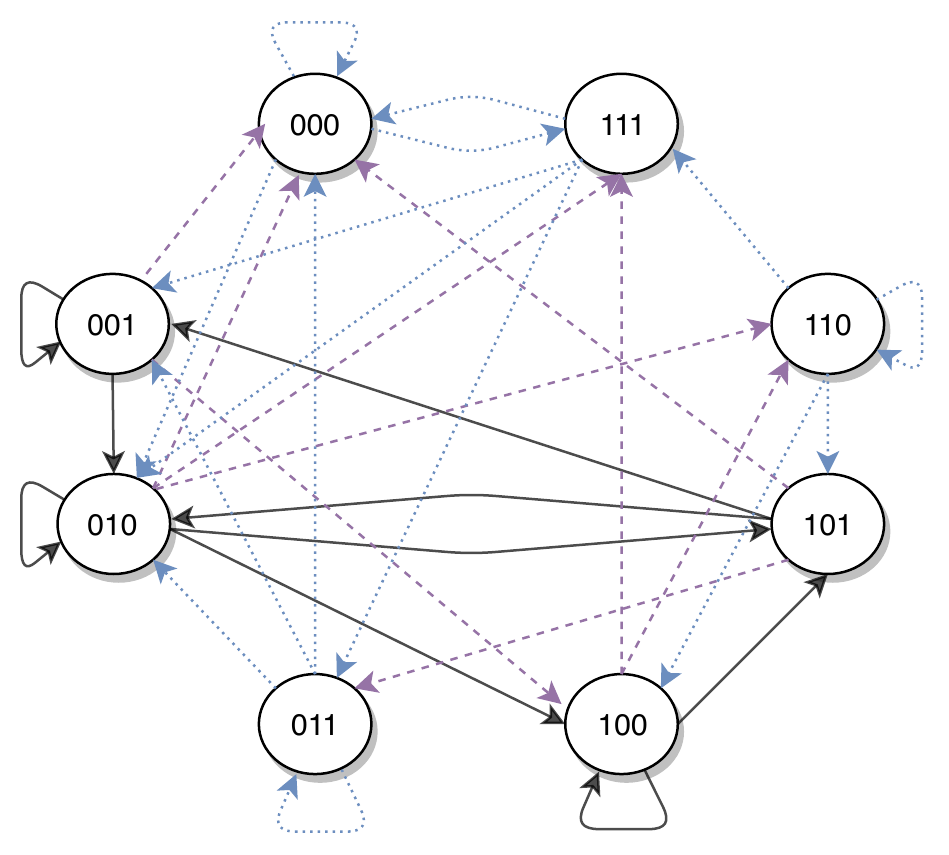}
\caption{ }
\label{fig:sub2}
\end{subfigure}
\caption{ An example for Steps \ref{step3} and \ref{step4} of Construction \ref{const:D} for $q=2$ and $k=l=1$. Figure (a) shows the graph $G^3$ obtained in Step~\ref{step3} of Construction \ref{const:D}. 
Figure (b) shows the graph $D$. The solid edges correspond to the edges in $G^3$, the dashed and dotted edges are added in Step \ref{step4} of Construction \ref{const:D}. The tails of the dashed edges are in $V(G^3)$ and the tails of the dotted edges are in $V(D)\setminus V(G^3)$.}
\label{fig:G3}
\end{figure*}

\subsection{Open problems}
The concept of recoverable systems gives rise to a group of open questions. First, the family of circulant graphs
is a subfamily of a larger class, namely, {\em transitive graphs} (when the automorphism group acts transitively on the set of vertices). Constructing recoverable systems that yield a bound on the capacity of storage codes for such graphs is an interesting open problem. Next, it is of interest to extend the construction of deterministic recoverable systems to the case $k,l>1$, yielding bounds on the capacity for the more general case than the $(1,1)$-recoverability considered above.

Several open questions arise for $(\epsilon,k,l)$-recoverable systems. Under our definition, the residual entropy of 
the group of symbols is constrained by $\epsilon$ for {\em every realization} of the neighborhood; see \eqref{eq:ekl}. It is of interest to analyze the capacity of recoverable systems when this requirement holds only on average, i.e., when the entropy of the symbol group is conditioned on the random variables $X_{[l]},X_{l+k+[l]}$ that form their neighborhood. Another open question
is providing characterization of the entropy maximizing measures without restricting the values of $\epsilon$.

\subsection*{Acknowledgments:} We are grateful to the reviewers of this submission for useful comments and corrections that have improved the presentation of the results. We also acknowledge a comment made by an anonymous reviewer of our conference paper \cite{EliBar2021} regarding a possible approach to the proof of Theorem \ref{th:final1}, summarized in the remark in the main text.

\appendix

\begin{example}\label{ex:cont}  (Refer to the discussion after Construction \ref{const:D}.) {\rm We start with the binary $(1,1)$-recoverable system given in Proposition~\ref{ex:1}. Let $Q$ be the binary 
alphabet, let $k=l=1$ and let $\epsilon=0.286$ which yields $\delta=0.05, \bar\delta=0.95$.
The following example shows the construction of the graph $D$ for the binary $(1,1)$-recoverable system with maximum
capacity given in Prop~\ref{ex:1}. The graph that presents this system is shown in Fig.~\ref{fig:G}.
\begin{figure}[h!]
		\centering\includegraphics[width=0.7\linewidth]{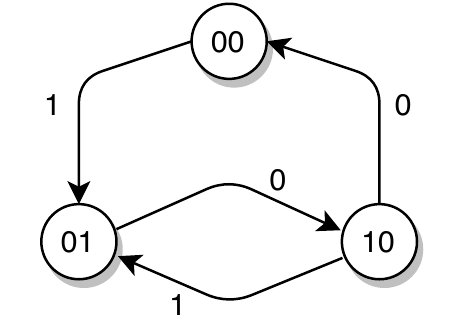}\\
	\caption{The graph $G'$ presenting a binary $(1,1)$-recoverable system with maximum capacity}\label{fig:G}
\end{figure}
\begin{figure*}[!t]
\begin{center}
\begin{align} 
\label{eq:pmatrixend}
    P=\begin{bmatrix}
    0.43 \delta & 0& 0.43 \bar\delta & 0& 0.245\bar\delta & 0.325\bar\delta&0.245\delta& 0.325\delta \\
    0.57\delta & 0.43 \bar\delta&0.57\bar\delta& 0.43\delta&0&0&0&0 \\
    0.43 \delta & 0& 0.43 \bar\delta & 0& 0.245\bar\delta & 0.325\bar\delta&0.245\delta& 0.325\delta \\
    0.57\delta & 0.43 \bar\delta&0.57\bar\delta& 0.43\delta&0&0&0&0 \\
    0&0&0&0& 0.43\bar\delta&0.57\bar\delta&0.43\delta& 0.57\delta \\
    0.57\delta& 0.43\bar\delta& 0.57\bar\delta& 0.43\delta &0&0&0&0 \\
    0&0&0&0& 0.43\bar\delta&0.57\bar\delta&0.43\delta& 0.57\delta \\
    0.57\delta& 0.43\bar\delta& 0.57\bar\delta& 0.43\delta &0&0&0&0
    \end{bmatrix}
\end{align}
\hrulefill
\begin{equation} 
\label{eq:pnuend}
\begin{split}
    p^{\nu}&=(0.411\delta,0.177\bar\delta ,0.411\bar\delta ,0.177\delta,0.177\bar\delta ,0.235\bar\delta ,0.177\delta,0.235\delta)\\
    &\approx (0.02,0.168,0.391,0.009,0.168,0.223,0.009,0.012)
    \end{split}
    \end{equation}
    \hrulefill
\end{center}
\end{figure*}

Applying Steps \ref{step1},\,\ref{step2} of  Construction~\ref{const:D}, we obtain the graph $G^3$ shown in Figure \ref{fig:sub1}. Now following Construction~\ref{const:markov},
we obtain $\mu=(0.177,0.411,0.177,0.235)$ and 
    $$
    P^{\mu}=\begin{bmatrix}
    0.43 & 0.57 & 0 & 0 \\
    0& 0.43 & 0.245 & 0.325 \\
    0 & 0 & 0.43 & 0.57 \\
    0.43 & 0.57 &0 &0 
    \end{bmatrix},
    $$
where we have ordered the vertices of $G^3$ by the increase of their numerical value.
    The graph $D$ is presented in Figure \ref{fig:sub2}. 
    The matrix $P$ is given in \eqref{eq:pmatrixend} at the top of the next page,  
    where the vertices are again ordered by the increase of their numerical value.
    The stationary distribution $p^{\nu}$ is given in \eqref{eq:pnuend} at the top of the next page.
For instance, we have
   $$
p_0=\frac{\nu([001])}{\nu([001])+\nu([011])}=\bar\delta
  $$
and 
\[p_1=\frac{\nu([001])}{\nu([001])+\nu([011])}=\delta.\]
}
\hfill\IEEEQED
\end{example}


\end{document}